\documentclass[amsmath,amssymb]{revtex4}
\usepackage{graphicx}
\usepackage{amscd,amsmath,amsthm,amsfonts,amssymb}
\usepackage[colorlinks,linkcolor=red,anchorcolor=blue,citecolor=green]{hyperref}
\usepackage{ytableau}
\usepackage{extarrows}

\newtheorem{thm}{Theorem}
\makeatletter \@addtoreset{equation}{section} \makeatother
\renewcommand{\theequation}{\arabic{section}.\arabic{equation}}
\def\[{\begin{equation}}
\def\]{\end{equation}}

\begin{document}
\title{Spatially-bounded rogue waves in the Davey-Stewartson I equation}
\author{
Bo Yang$^{1}$, Jianke Yang$^{2}$}
\address{$^{1}$ School of Mathematics and Statistics, Ningbo University, Ningbo 315211, China\\
$^{2}$ Department of Mathematics and Statistics, University of Vermont, Burlington, VT 05401, U.S.A}
\begin{abstract}
Spatially-bounded rogue waves, i.e., rogue waves that arise in a limited region of a multi-dimensional space, are interesting and important from both theoretical and applied points of view. In this paper, we determine spatially-bounded rogue waves in the Davey-Stewartson I equation. We show that these rogue waves can be obtained when a single or multiple internal parameters in the higher-order rational solution of the Davey-Stewartson I equation are real and large, and the order-index vector of this higher-order rational solution has even length and comprises pairs of the form $(2n, 2n+1)$, where $n$ is a positive integer. Under these conditions and another nondegeneracy condition on the root curve of a certain double-real-variable polynomial, the higher-order rational solution will exhibit spatially-bounded rogue waves that arise from a uniform background with some time-varying lumps on it, reach high amplitude in limited space, and then disappear into the same background again. The crests of these rogue waves form a single or multiple closed curves that are generically disconnected from each other on the spatial plane, and are analytically predicted by the root curve mentioned above. We also derive uniformly-valid asymptotic approximations for these spatially-bounded rogue waves in the large-parameter regime. Near the crests of these rogue waves, these asymptotic approximations reduce to simple expressions. Our asymptotic approximations of these rogue waves are compared to true solutions and good agreement is demonstrated.
\end{abstract}
\maketitle

\section{Introduction}
Rogue waves are spontaneous and extreme wave excitations that occur most famously in water but also in other physical systems \cite{PeliBook,Solli_Nature,Wabnitz_book,YangYangbook}. These waves have several key characteristics. First, they appear and then disappear quickly without obvious warning. Second, when they arise, they reach amplitudes much higher than their original amplitudes (at least two times or more). Thirdly, the energy of these waves is highly concentrated in space. Due to their elusive and extreme nature and potential damage, a lot of experimental work has been done on rogue waves in diverse physical systems, such as water tanks \cite{Tank1,Tank2, Tank3, Tank4}, optical fibers \cite{Solli_Nature,Wabnitz_book,Fiber1,Fiber1b}, plasma \cite{Plasma}, Bose-Einstein condensates \cite{RogueBEC}, acoustics \cite{Acoustics}, and so on.

Theoretically, many such physical rogue waves can be described by rational solutions in certain integrable equations, such as the nonlinear Schr\"odinger (NLS) equation, the Manakov system, and so on. There are two reasons for this. One reason is that these integrable equations are known to govern those physical processes \cite{BenneyNewell,Menyuk}. The other reason is that those rational solutions often share the key characteristics of physical rogue waves mentioned above and can thus be called theoretical rogue waves \cite{Peregrine,AAS2009,DGKM2010,KAAN2011,GLML2012,OhtaJY2012,BDCW2012,ManakovDark,LingGuoZhaoCNLS2014,Chen_Shihua2015,ZhaoGuoLingCNLS2016}. Importantly, these theoretical rogue waves in integrable systems can be explicitly derived using the integrable theory. These explicit rogue wave solutions are mathematically clean, and studies of them can reveal the most fundamental physical mechanisms and most important dynamical features of physical rogue waves, thus contributing to our understanding and prediction of rogue events.

In this paper, we consider theoretical rogue waves in the Davey-Stewartson-I (DSI) equation. This equation governs the evolution of a two-dimensional surface wave packet in shallow water under strong surface tension and is integrable \cite{Benney_Roskes,Davey_Stewartson,Ablowitz_book}. Rogue waves in this equation have been studied in \cite{OhtaYangDSI,YangCurve2024,HeDSI_2021}. In \cite{OhtaYangDSI,YangCurve2024}, rational solutions were derived, and it was shown that those rational solutions could exhibit various types of rogue waves such as line rogue waves, multi-line rogue waves, and rogue waves whose crests form complex curves on the spatial plane. These rogue waves arise from the uniform background (with possibly some lumps on it), reach higher amplitude and then retreat to the same background again. Most of these reported rogue waves are spatially unbounded, meaning that they appear in an unbounded spatial region. In \cite{HeDSI_2021}, spatially-bounded rogue waves on a background of dark solitons were reported. These rogue waves in \cite{HeDSI_2021} are semi-rational, i.e., their analytical expressions contain both exponential and rational terms.

Spatially-bounded rogue waves that arise in a limited region of a multi-dimensional space are physically important, because such rogue waves can be generated by perturbations to the background in a spatially limited region, which is more manageable in an experimental setting. Thus, we are asking the following question: what spatially-bounded rogue waves can appear in DSI that arise from an almost-uniform background (rather than from a dark-soliton background)? In our earlier work \cite{YangCurve2024}, we have shown that such spatially-bounded rogue waves do exist in the family of higher-order rational solutions of DSI when an internal parameter in those solutions is real and large. But more examples of such higher-order rational solutions in \cite{YangCurve2024} exhibit only spatially-unbounded rogue waves. So it is unclear yet which of those countless higher-order rational solutions admit spatially-bounded rogue waves. In addition, asymptotic approximations derived in \cite{YangCurve2024} for those rogue waves were limited to certain spatial regions and were not valid at places such as horizontal edges of rogue waves. Thus, a full asymptotic description of spatially-bounded rogue waves is still missing.

In this paper, we address these open questions. We first show that spatially-bounded rogue waves in DSI can be obtained when a single or multiple internal parameters in the higher-order rational solution of the DSI equation are real and large, and the order-index vector of this higher-order rational solution has even length and comprises pairs of the form $(2n, 2n+1)$, where $n$ is a positive integer. Under these conditions and another nondegeneracy condition on the root curve of a certain double-real-variable polynomial, the higher-order rational solution of DSI will exhibit a spatially-bounded rogue wave that arises from a uniform background with some time-varying lumps on it, reach high amplitude in limited space, and then disappear into the same background again. The crests of this spatially-bounded rogue wave form a single or multiple closed curves that are generically disconnected from each other on the spatial plane. Next, we analytically derive uniformly-valid asymptotic approximations for this spatially-bounded rogue wave. We show that the crests of this rogue wave are predicted by the root curve mentioned above, and near these crests the rogue solution has simple asymptotic expressions.  Our asymptotic approximations of these rogue waves are compared to true solutions and good agreement is demonstrated.

\section{Preliminaries}
Evolution of a two-dimensional wave packet on water of finite depth is governed by the Benney-Roskes-Davey-Stewartson equation \cite{Benney_Roskes,Davey_Stewartson,Ablowitz_book}. In the shallow water limit, this equation is integrable (see Ref. \cite{Ablowitz_book} and the references therein). This integrable equation is sometimes just called the DS equation in the literature. The DS equation is divided into two types, DSI and DSII, which correspond to the strong surface tension and weak surface tension, respectively \cite{Ablowitz_book}.

The DSI equation is
\begin{equation} \label{DS}
\left.
\begin{array}{l}
\textrm{i}A_t=A_{xx} + A_{yy}+ (\epsilon|A|^2-2 Q)A,
\\[5pt]
Q_{xx}-Q_{yy}=\epsilon (|A|^2)_{xx},
\end{array} \right\}
\end{equation}
where $\epsilon=\pm 1 $ is the sign of nonlinearity. In the two-dimensional surface water wave context where it was first derived \cite{Davey_Stewartson,Ablowitz_book}, $A$ is the complex envelope function of the surface wave packet that propagates along the $x$ direction, $Q$ is the $x$-direction velocity of the mean flow, and $\epsilon=1$.

Rational solutions in DSI have been derived in \cite{OhtaYangDSI} and simplified in \cite{YangCurve2024}. Those rational solutions contain various types of solutions, such as multi-line rogue waves and higher-order rational solutions, depending on whether the spectral parameters in them are different or the same. In addition, those solutions contain many free internal complex parameters. In this article, we consider the higher-order rational solutions with their internal parameters under certain restrictions. Explicit expressions of these restricted higher-order rational solutions have been presented in \cite{YangCurve2024} and are quoted below.

First, we introduce elementary Schur polynomials $S_n(\mbox{\boldmath $x$})$ with $ \emph{\textbf{x}}=\left( x_{1}, x_{2}, \ldots \right)$, which are defined by the generating function
\begin{equation}\label{Elemgenefunc}
\sum_{n=0}^{\infty}S_n(\mbox{\boldmath $x$}) \epsilon^n
=\exp\left(\sum_{j=1}^{\infty}x_j \epsilon^j\right),
\end{equation}
and $S_n\equiv 0$ if $n<0$. Then, these restricted higher-order rational solutions are given by
\[ \label{DSAQ}
  A(x,y,t)= \sqrt{2}\frac{\tau_1}{\tau_0},  \qquad
  Q(x,y,t)= 1-2 \epsilon \left( \log \tau_0 \right)_{xx},
\]
where
\[ \label{deftaunk}
\tau_{k}=
\det_{
\begin{subarray}{l}
1\leq i, j \leq N
\end{subarray}
}
\left(
\begin{array}{c}
m_{i,j}^{(k)}
\end{array}
\right),
\]
$N$ is a positive integer,
the matrix elements $m_{i,j}^{(k)}$ are defined by
\[ \label{Schmatrimnij}
m_{i,j}^{(k)}=\sum_{\nu=0}^{\min(n_{i}, n_{j})}\frac{1}{4^\nu} \hspace{0.06cm} S_{n_{i}-\nu}[\textbf{\emph{x}}^{+}(k) +\nu \textbf{\emph{s}} ] \hspace{0.06cm} S_{n_{j}-\nu}[\textbf{\emph{x}}^{-}(k) + \nu \textbf{\emph{s}}],
\]
$(n_1, n_2, \dots, n_N)$ is an order-index vector with each $n_i$ a free positive integer, $n_1<n_2<\cdots <n_N$, vectors $\textbf{\emph{x}}^{\pm}(k)=\left( x_{1}^{\pm}, x_{2}^{\pm},\cdots \right)$ are defined by
\begin{eqnarray}
&&x_{j}^{+}(k)= \frac{(-1)^j}{j!2p} \epsilon(x-y) - \frac{(-2)^{j-1}}{j!p^2} \textrm{i}t + \frac{1}{j!2} p (x+y) - \frac{2^{j-1}}{j!} p^2 \textrm{i}t  + k\delta_{j, 1}+a_j,    \label{xrijplus}\\
&&x_{j}^{-}(k)= \frac{(-1)^j}{j!2p} \epsilon(x-y) + \frac{(-2)^{j-1}}{j!p^2} \textrm{i}t + \frac{1}{j!2} p (x+y) + \frac{2^{j-1}}{j!} p^2 \textrm{i}t  - k \delta_{j,1}+a_j^*,  \label{xrijminus}
\end{eqnarray}
$p$ is a free real nonzero constant, $\delta_{j, 1}$ is the Kronecker delta function which is equal to 1 when $j=1$ and 0 otherwise, the asterisk `*' represents complex conjugation, $a_{1}, a_2, \dots, a_{n_N}$ are free complex constants, and $\textbf{\emph{s}}=(0, s_2, 0, s_4, \cdots)$ are coefficients from the expansion
\[   \label{schurcoeffsr}
\ln \left(\frac{2}{\kappa} \tanh \frac{\kappa}{2}\right) = \sum_{j=1}^{\infty}s_{j} \kappa^j.
\]

We would like to point out that in our previous work \cite{YangCurve2024}, the above solutions were called ``higher-order rogue wave solutions''. We now think that name is not accurate, because some of these solutions are not rogue wave solutions at all as they do not contain wave components that appear and disappear quickly without warning. One such example can be found in the lower row of Fig.~\ref{f:fig6} in the later text. For this reason,  in this article we call these solutions ``higher-order rational solutions'' and reserve the name ``rogue wave'' to only wave components in these solutions that appear and then disappear quickly. Hopefully, this can reduce confusion.

Under the variable transformation of $Q \rightarrow Q + \epsilon |A|^2,\ x \leftrightarrow y$, and $\epsilon \rightarrow - \epsilon $, the DSI equation (\ref{DS}) is invariant. In addition, we have seen in \cite{YangCurve2024} that the above rational solution with $p\ne 1$ would be a skewed version of the $p=1$ solution. Thus, we will set
\[  \label{epsipa1}
\epsilon=1, \quad p=1
\]
in this article without loss of generality. Under these parameters, the $\textbf{\emph{x}}^{\pm}(k)$ vectors in Eqs.~(\ref{xrijplus})-(\ref{xrijminus}) reduce to
\begin{eqnarray}
&&x_{j}^{+}(k)= \frac{1+(-1)^j}{j!2} x + \frac{1-(-1)^j}{j!2} y-
\frac{2^{j}-(-2)^{j}}{j!2} \textrm{i}t  + k\delta_{j, 1}+a_j,    \label{xrijplus2}\\
&&x_{j}^{-}(k)= \frac{1+(-1)^j}{j!2} x + \frac{1-(-1)^j}{j!2} y+
\frac{2^{j}-(-2)^{j}}{j!2} \textrm{i}t - k \delta_{j,1}+a_j^*.  \label{xrijminus2}
\end{eqnarray}
Notice that $x_1^{+}=y- 2 \textrm{i} t + k+a_1$ and $x_1^{-}=y+2 \textrm{i} t - k+a_1^*$. Thus, we normalize $a_1=0$ through a $(y, t)$ coordinate shift. Also notice that  $x_2^+=\frac{1}{2}x +a_2$ and  $x_2^-=\frac{1}{2}x +a_2^*$. Thus, we normalize $\Re(a_2)=0$ through an $x$-coordinate shift, where $\Re$ represents the real part of a complex parameter. We also denote the parameter vector $\textbf{\emph{a}}\equiv (0, a_2, \dots, a_{n_N})$ and order-index vector $\Lambda\equiv (n_1, n_2, \dots, n_N)$.

The simplest rational solution (\ref{DSAQ}) is obtained when we set $N=n_1=1$, in which case
\[ \label{1strogue}
A(x, y, t)=\sqrt{2} \left[1+\frac{16{\rm{i}}t-4}{4y^2+16t^2+1}\right].
\]
This is a line rogue wave centered along the $x$-axis and exhibiting the Peregrine profile \cite{Peregrine} along the $y$-direction. It rises from the constant background of amplitude $\sqrt{2}$, reaches higher amplitude of three times the background on a horizontal line crest, and then retreats to the same constant background again. This line rogue wave has been reported in \cite{OhtaYangDSI}.

\begin{figure}[htb]
\begin{center}
\includegraphics[scale=0.27, bb=1440 0 300 570]{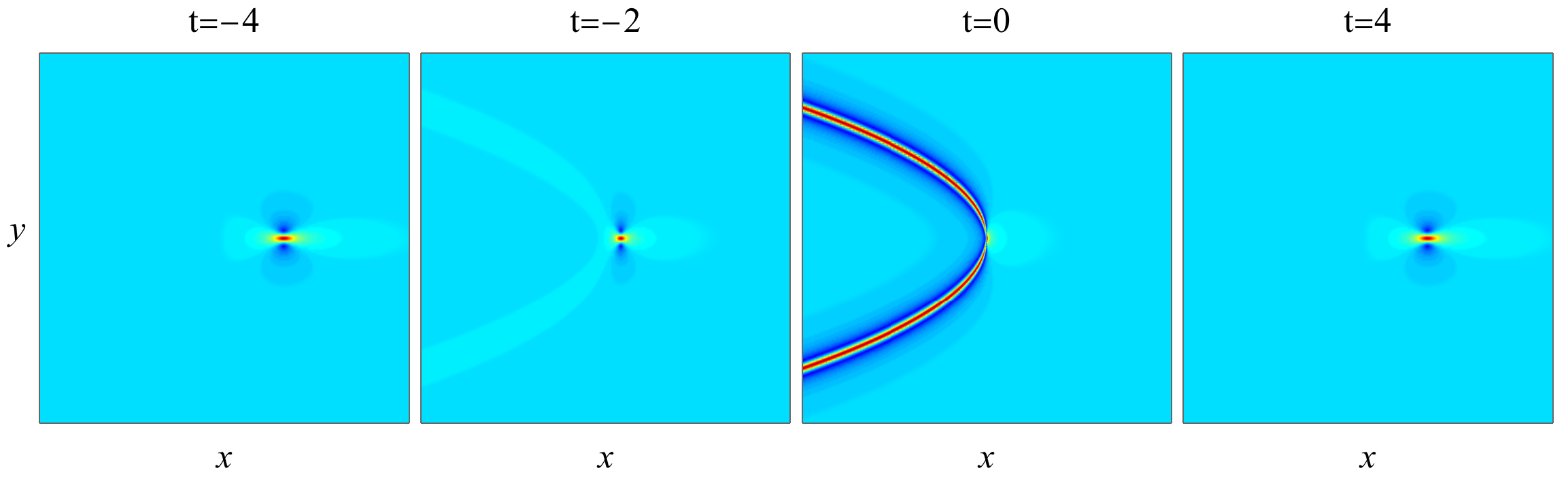}
\caption{The second-order rational solution $|A(x, y, t)|$ from Eqs.~(\ref{2ndrogue})-(\ref{2ndroguetauk}) with $a_2=0$ at four time values of $t=-4, -2, 0$ and 4. In all panels, $-200\le x\le 200$ and $-20\le y\le 20$.} \label{f:fig1}
\end{center}
\end{figure}

The next simplest rational solution (\ref{DSAQ}) is obtained when we set $N=1$ and $n_1=2$. In this case,
\[  \label{2ndrogue}
A(x, y, t)=\sqrt{2}\frac{\tau_1}{\tau_0},
\]
where
\[ \label{2ndroguetauk}
\tau_k=\left[\frac{1}{2}x+a_2+\frac{1}{2} (y-2{\rm{i}}t+k)^2\right] \left[\frac{1}{2}x+a_2^*+\frac{1}{2} (y+2{\rm{i}}t-k)^2\right]+\frac{1}{4}(y-2{\rm{i}}t+k)(y+2{\rm{i}}t-k)+\frac{1}{16}.
\]
We call this solution the second-order rational solution, and it contains a free purely-imaginary constant $a_2$. If we take $a_2=0$, this $A(x, y, t)$ solution is illustrated in Fig.~\ref{f:fig1}. It starts from a lump (see the $t=-4$ panel). As time increases, this lump moves to the left and shrinks in size horizontally, and its peak amplitude remains roughly three times the background. Simultaneously, a rogue wave in the shape of a left-facing parabola rises from the constant background (see the $t=-2$ and 0 panels). This parabola-shaped rogue wave reaches peak amplitude about three times the constant background on its parabola crest at $t=0$. Meanwhile, the lump becomes very thin horizontally. When time increases further from 0, the process is reversed, i.e., the parabola-shaped rogue wave disappears, and simultaneously the lump moves to the right and expands in size horizontally, see the $t=4$ panel. This rogue solution was first reported in \cite{OhtaYangDSI}. An interesting feature of this solution is that in addition to the parabola-shaped rogue part, it also contains a time-varying lump on the constant background that does not disappear at large time $|t|$. So, we can say this is a rogue wave that sits on a constant background with a time-varying lump.

Notice that the DSI equation (\ref{DS}) is invariant when $x$ is switched to $-x$. Thus, when we replace $x$ by $-x$ in Eq.~(\ref{2ndroguetauk}), we would get another second-order rational solution (\ref{2ndrogue}) whose $\tau_k$ function is
\[ \label{2ndroguetauk2}
\tau_k=\left[-\frac{1}{2}x+a_2+\frac{1}{2} (y-2{\rm{i}}t+k)^2\right] \left[-\frac{1}{2}x+a_2^*+\frac{1}{2} (y+2{\rm{i}}t-k)^2\right]+\frac{1}{4}(y-2{\rm{i}}t+k)(y+2{\rm{i}}t-k)+\frac{1}{16},
\]
where $a_2$ is a free purely-imaginary constant (recall that $\Re(a_2)$ has been normalized to zero through an $x$-coordinate shift). When $a_2=0$, the graph of this associated rogue solution is a horizontal reflection of that in Fig.~\ref{f:fig1}.

In the line rogue wave (\ref{1strogue}), the rogue crest is a straight horizontal line. In the second-order rational solutions (\ref{2ndrogue})-(\ref{2ndroguetauk2}), the rogue crest (that appears and then disappears) is a parabola. In both cases, the rogue waves that arise are unbounded in space. In this article, we are looking for rogue waves that are bounded in space. Such rogue waves do exist in the class of higher-order rational solutions (\ref{DSAQ}) under certain conditions, and we will reveal those conditions in the next section.

\section{Spatially-bounded rogue waves}

Spatially-bounded rogue waves are significant both theoretically and practically. To generate such rogue waves in an experimental setting, an important question is what initial conditions need to be prepared. The best way to answer that question is to analytically determine such rogue waves and then use those analytical solutions to prepare initial conditions, as was done in many past rogue wave experiments \cite{Tank1,Tank2, Tank3, Tank4, Fiber1,Fiber1b,Plasma,RogueBEC,Acoustics}.

In this section, we analytically study which of the higher-order rational solutions (\ref{DSAQ}) exhibit spatially-bounded rogue waves. It turns out that the answer to this question is simple if internal parameters in these higher-order rational solutions are such that $a_2=O(1)$, while the other internal parameters $(a_3, a_4, \dots a_{n_N})$ are real and large in the form
\[ \label{acond}
a_j=\kappa_{j-2} R^j, \quad 3\le j\le n_N,
\]
where $R\gg 1$ is a large positive constant, and $(\kappa_1, \kappa_2, \dots, \kappa_{n_N-2})$ are arbitrary $O(1)$ real constants not being all zero.

To present our answer to this case, we need to introduce a class of polynomials $\mathcal{P}_{\Lambda}(z_1, z_2)$ in two real variables $(z_1, z_2)$, which will arise naturally in our later analysis. These polynomials are defined as
\begin{equation} \label{DoubleRealPolydef}
\mathcal{P}_{\Lambda}(z_1, z_2) =\det_{1\le i\le N}\left[
         H_{n_i}(z_1, z_2), \hspace{0.05cm} H_{n_i-1}(z_1, z_2), \hspace{0.05cm} \cdots, \hspace{0.05cm}, H_{n_i-N+1}(z_1, z_2) \right],
\end{equation}
where $H_{n}(z_1, z_2)$ are Schur polynomials generated by the expansion
\begin{equation}\label{AMthetak}
\sum_{n=0}^{\infty} H_n(z_1, z_2) \epsilon^n =\exp\left( z_2 \epsilon + z_1 \epsilon^{2} +
\sum_{j=1}^{\infty} \kappa_{j} \epsilon^{j+2}\right),
\end{equation}
with $H_{n}(z_1, z_2)\equiv 0$ if $n<0$, $\Lambda=(n_1, n_2, \dots, n_N)$ is an order-index vector which we have seen in the previous section, and parameters $\kappa_j \hspace{0.04cm} (j\ge 1)$ are real constants. Notice that $H_{n}(z_1, z_2)=S_{n}(z_2, z_1, \kappa_1, \kappa_2, \dots)$ in view of their definitions in Eqs.~(\ref{Elemgenefunc}) and (\ref{AMthetak}). These $ \mathcal{P}_{\Lambda}(z_1, z_2)$ polynomials generalize those introduced in \cite{YangCurve2024} where all $\{\kappa_j\}$ parameters were zero except for one of them. Notice that $\mathcal{P}_{\Lambda}(z_1, z_2)$ depends on parameters $(\kappa_1, \dots, \kappa_{n_N-2})$ since its highest matrix-element polynomial $H_{n_N}(z_1, z_2)$ depends on these parameters. This determinant in (\ref{DoubleRealPolydef}) is a Wronskian in $z_2$ since we can see from Eq.~(\ref{AMthetak}) that $(\partial/\partial z_2)H_{n}(z_1, z_2)=H_{n-1}(z_1, z_2)$.

By setting
\[ \label{Pmz1z2}
\mathcal{P}_{\Lambda}(z_1, z_2)=0
\]
for real values of $(z_1, z_2)$, we get solution sets in the $(z_1, z_2)$ plane. These solution sets are generically a collection of curves, which we will call as root curves. Examples of such root curves will be seen in Fig.~\ref{f:fig3} shortly.
But in some cases, these root curves are degenerate, meaning that they comprise only isolated points in the $(z_1, z_2)$ plane. For example, if $\Lambda=(4,5)$ and $(\kappa_1, \kappa_2, \kappa_3)=(0, -1, 0)$, then
\[ \label{Pdege}
\mathcal{P}_{\Lambda}(z_1, z_2)=\frac{1}{2880} \left\{ 720(z_1^2-2)^2+z_2^4\left[(z_2^2+8z_1)^2+56z_1^2+240\right]\right\}.
\]
Since this function is a sum of squares, it is easy to see that the solution set $(z_1, z_2)$ of Eq.~(\ref{Pmz1z2})  comprises only two isolated points $(\pm \sqrt{2}, 0)$. Thus, root curves are degenerate here. It might be even possible for the solution set of Eq.~(\ref{Pmz1z2}) to be entirely empty, even though we have not found such examples yet.

Now, we are ready to present our answer to the question of which higher-order rational solutions (\ref{DSAQ}) exhibit spatially-bounded rogue waves if their internal parameters meet the conditions (\ref{acond}).

\begin{thm} \label{Theorem1}
The higher-order rational solution (\ref{DSAQ}) under parameter conditions (\ref{acond}) would exhibit spatially-bounded rogue waves if it satisfies the following two conditions:
\begin{enumerate}
\item $N$ is even, and the order-index vector $(n_1, n_2, \dots, n_N)$ is a concatenation of pairs of the form $(2n, 2n+1)$, where $n$ is a positive integer; and
\item root curves of Eq.~(\ref{Pmz1z2}) are not degenerate or empty.
\end{enumerate}
\end{thm}

According to the first condition of this theorem, the higher-order rational solution (\ref{DSAQ}) under parameter conditions (\ref{acond}) could exhibit spatially-bounded rogue waves if its order-index vector $(n_1, n_2, \dots, n_N)$ is of a form such as $(2, 3)$, $(2, 3, 4, 5)$, $(4, 5, 8, 9, 12, 13)$, etc. The line rogue wave (\ref{1strogue}) and the second-order rational solution (\ref{2ndrogue})-(\ref{2ndroguetauk}) have order indices of $(1)$ and $(2)$ respectively, which do not meet the above order-index condition. Thus, it is not surprising that their rogue waves are not spatially-bounded (see Fig.~\ref{f:fig1}).

\begin{figure}[htb]
\begin{center}
\includegraphics[scale=0.27, bb=1440 0 300 570]{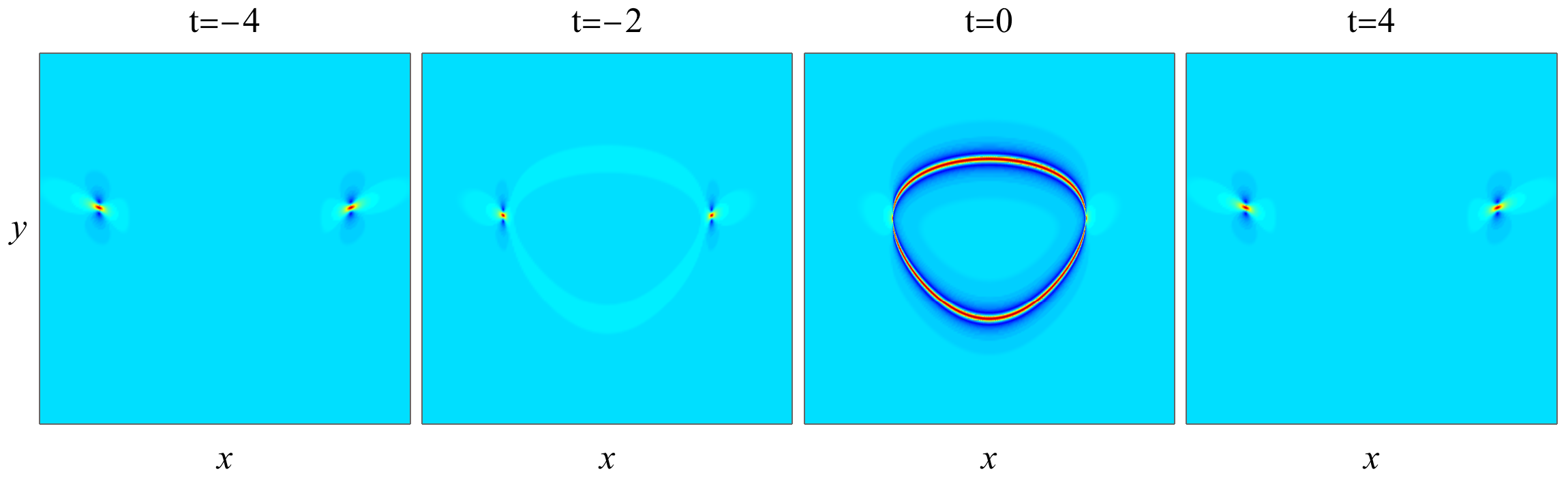}
\caption{The simplest spatially-bounded rogue wave $|A(x, y, t)|$ from Eqs.~(\ref{simpleLRogue})-(\ref{simpleLRogue2}) with $a_3=1000$ at four time values of $t=-4, -2, 0$ and 4. In all panels, $-400\le x\le 400$, and $-15\le y\le 38$.} \label{f:fig2}
\end{center}
\end{figure}

Based on Theorem~\ref{Theorem1}, the simplest rational solution (\ref{DSAQ}) that exhibits a spatially-bounded rogue wave under parameter conditions (\ref{acond})  would have $N=2$ and order index $(n_1, n_2)=(2, 3)$. This rational solution contains two internal parameters $a_2$ and $a_3$, where $a_2$ is $O(1)$ purely-imaginary and $a_3=\kappa_1 R^3$, with $\kappa_1$ being $O(1)$ real and $R$ a large positive constant (basically, $a_3$ is just a real constant with large magnitude). It is easy to check that the root curve of Eq.~(\ref{Pmz1z2}) for
$\Lambda=(2, 3)$ and $\kappa_1\ne 0$ is regular (i.e., not degenerate or empty). Thus, Theorem~\ref{Theorem1} predicts that the rational solution (\ref{DSAQ}) with order index $(n_1, n_2)=(2, 3)$ and large real constant $a_3$ (in magnitude) would exhibit a spatially-bounded rogue wave. To check this prediction, let us take $a_2=0$ for simplicity. In this case, the analytical expression for $A(x, y, t)$ of this rational solution is
\[ \label{simpleLRogue}
A(x, y, t)=\sqrt{2}\left(1+\frac{G_1+{\rm{i}}G_2}{F_0}\right),
\]
where
\begin{eqnarray}
&& F_0=2048 t^8+512 \left(4 y^2+21\right) t^6+32 \left(24 y^4-28 y^2+24 x^2-4 x+371\right) t^4   \nonumber \\
&&\hspace{0.76cm}   +16 \left[8 y^6-58 y^4+103 y^2+x^2 \left(102-72 y^2\right)+x \left(12 y^2-17\right)+35\right]
   t^2 +8 y^8-8 y^6  \nonumber \\
&&\hspace{0.76cm}  +72 x^4+48 x^2 y^4-8 x y^4+54 y^4-24 x^3+38 x^2-120 x^2 y^2+20 x y^2+32 y^2-6 x +5 \nonumber \\
&&\hspace{0.76cm}  +288 a_3^2\left(16 t^2+4 y^2+1\right)
-96 a_3 y \left[-96 t^4-4 \left(4 y^2+11\right) t^2+2 y^4+6 x^2-3 y^2-x-1\right] ,
\end{eqnarray}
\begin{eqnarray}
&& G_1=-4 \left[3584 t^6+480 \left(4 y^2+11\right) t^4+12 \left(24 y^4-4 y^2+24 x^2-4 x+25\right) t^2+8 y^6+26 y^4 \right.  \nonumber \\
&& \hspace{1.3cm}  \left.  -6 x^2-72 x^2 y^2+12 x y^2+13 y^2+x+288 a_3^2+24 a_3 y \left(144 t^2+4
   y^2+5\right) \right]-6,
\end{eqnarray}
and
\begin{eqnarray}
&&G_2=16 t \left[512 t^6+96 \left(4 y^2+9\right) t^4+4 \left(24 y^4-76 y^2+24 x^2-4 x+15\right) t^2+8 y^6-34 y^4  \right.  \nonumber \\
&&\hspace{1.3cm}   \left. +30 x^2-72 x^2 y^2+12 x y^2-29 y^2-5 x-2+288 a_3^2+24 a_3 y \left(48 t^2+4
   y^2-1\right) \right].   \label{simpleLRogue2}
\end{eqnarray}
This solution with $a_3=1000$ at four time values of $-4$, $-2$, 0 and 4 is plotted in Fig.~\ref{f:fig2}. We see that in the $t=-4$ panel, the solution shows two lumps on the constant background. These two lumps are similar to the lump in the $t=-4$ panel of the second-order rational solution in Fig.~\ref{f:fig1} and are moving toward each other and shrinking in their horizontal sizes. In the $t=-2$ and 0 panels, a spatially bounded rogue wave in the shape of a ring appears between these lumps. Peak amplitudes of this rogue ring are about three times the constant background. In the $t=4$ panel, this rogue ring disappears, and the two lumps reverse their directions and move away from each other.  These graphs confirm that a spatially-bounded rogue wave indeed appears in the solution (\ref{simpleLRogue})-(\ref{simpleLRogue2}).

It is helpful to view this solution in Fig.~\ref{f:fig2} as built by connecting a second-order rational solution (\ref{2ndroguetauk}) with its $x$-reversed counterpart (\ref{2ndroguetauk2}) under certain $(x, y)$ positional shifts. Under this view, the bounded rogue ring in Fig.~\ref{f:fig2} is formed by linking the right-facing rogue parabola of Eq.~(\ref{2ndroguetauk2}) with the left-facing rogue parabola of Eq.~(\ref{2ndroguetauk}). This view is justified because, as we will show later in Sec.~\ref{sec:edge}, the left and right edges of the rogue ring in the $t=-2$ and 0 panels of Fig.~\ref{f:fig2} at large $a_3$ are asymptotically indeed second-order rational solutions (\ref{2ndroguetauk2}) and (\ref{2ndroguetauk}) under $(x, y)$ positional shifts respectively.

We should point out that the function (\ref{simpleLRogue}) is a valid DSI solution for any real $a_3$ value, not just for large $|a_3|$. But this solution exhibits a spatially-bounded rogue wave only when $|a_3|$ is large. Indeed, when we plot this solution for $O(1)$ values of $a_3$, we do not see spatially-bounded rogue waves (or any rogue waves at all). For example, when $a_3=0$, the graphs of this solution at time values of $t=-4, -2, 0$ and 4 resemble the right halves (or left halves) of the solutions shown in the lower row of Fig.~\ref{f:fig6} (with proper $(x, y)$-positional shifts), which clearly do not contain any rogue waves.

\subsection{Numerical confirmation of Theorem~\ref{Theorem1}}

In this subsection, we will use more examples to confirm Theorem~\ref{Theorem1}.

\begin{figure}[htb]
\begin{center}
\includegraphics[scale=0.26, bb=1530 0 300 570]{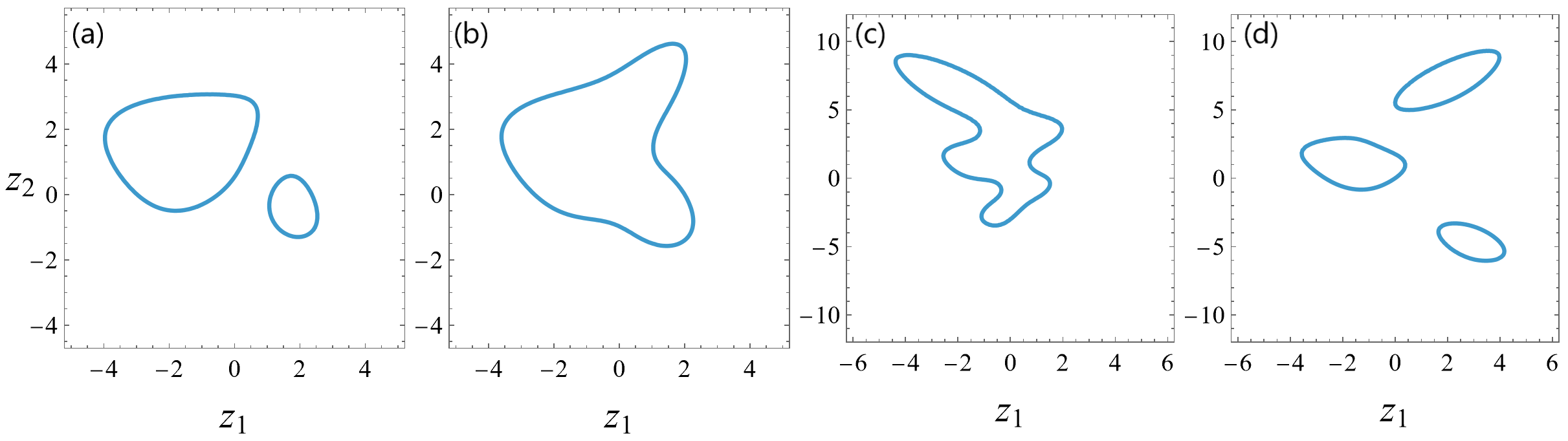}
\caption{Root curves of Eq.~(\ref{Pmz1z2}) for parameters (\ref{paraFig3a})-(\ref{paraFig4b}) of Examples 1-4 respectively. } \label{f:fig3}
\end{center}
\end{figure}

\begin{figure}[htb]
\begin{center}
\includegraphics[scale=0.27, bb=1440 0 300 570]{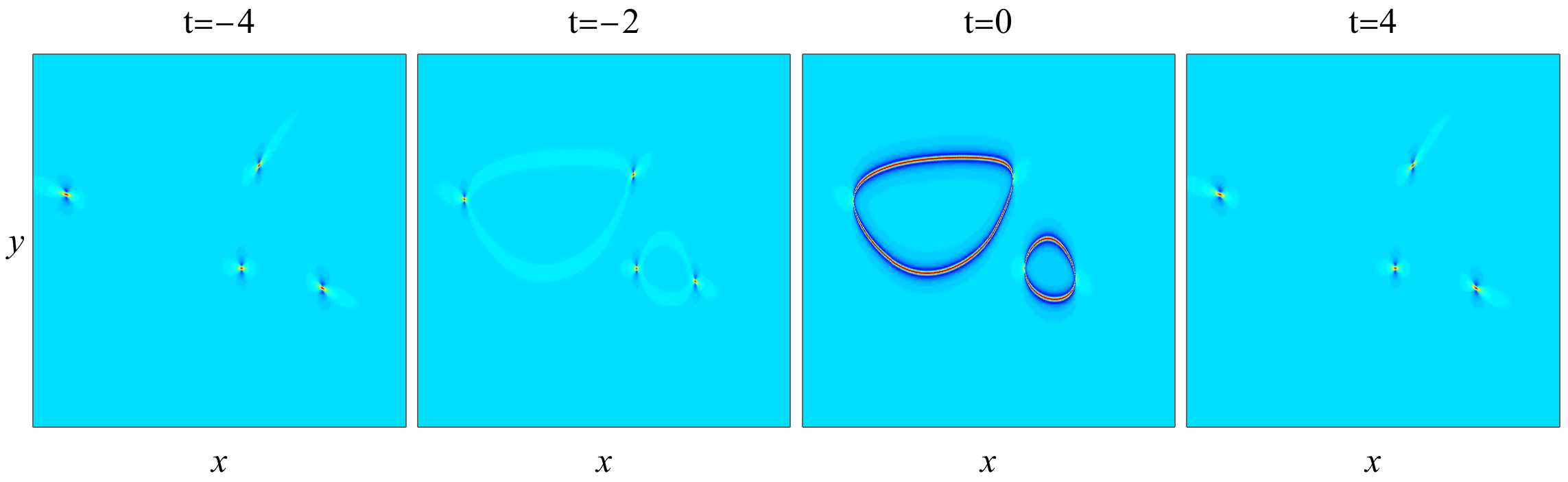}

\includegraphics[scale=0.27, bb=1440 0 300 570]{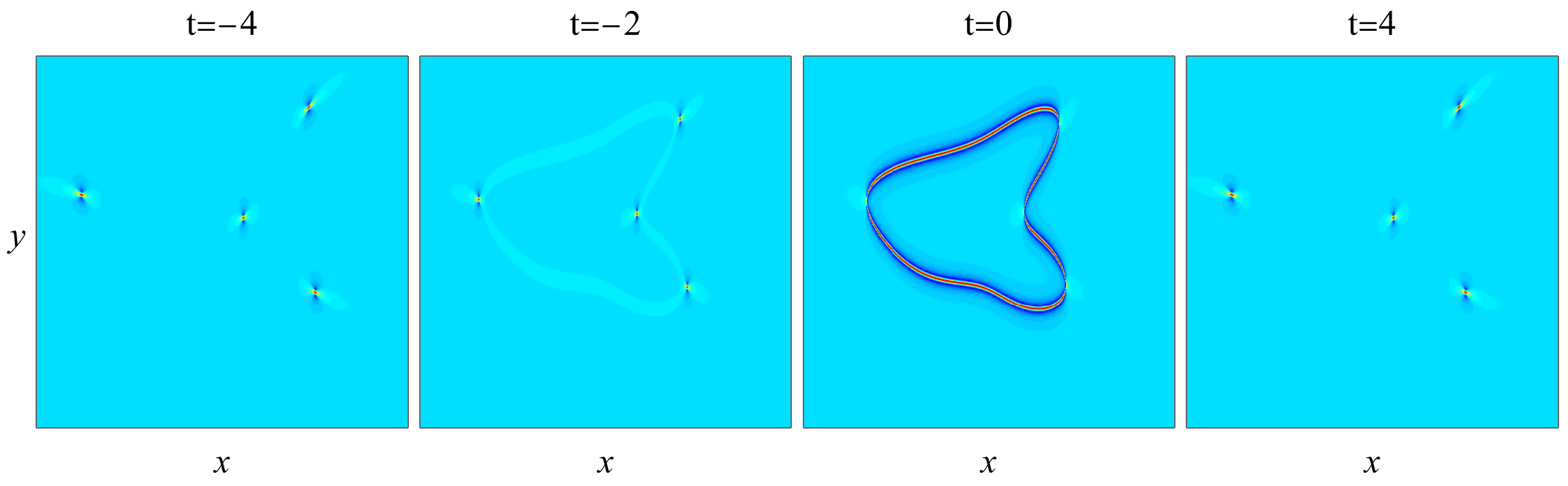}
\caption{Two spatially-bounded rogue waves $|A(x, y, t)|$ (Examples 1 and 2) to confirm Theorem~\ref{Theorem1}. These graphs are obtained by plotting rational solutions (\ref{DSAQ}) with parameter values (\ref{paraFig3a}) (upper row) and (\ref{paraFig3b}) (lower row) at four time values of $t=-4, -2, 0$ and 4. In all panels, $-700\le x\le 700$ and $-42\le y\le 50$.} \label{f:fig4}
\end{center}
\end{figure}

\textbf{Example 1. } In our first example, we choose parameters
\[ \label{paraFig3a}
N=4, \quad \Lambda=(2, 3, 4, 5), \quad R=8, \quad a_2=0, \quad a_3=R^3, \quad a_4=2R^4, \quad a_5=3R^5
\]
in the higher-order rational solution (\ref{DSAQ}). This order-index vector $\Lambda$ meets the first condition of Theorem~\ref{Theorem1}. To check the second condition of that theorem, we show in Fig.~\ref{f:fig3}(a) the root curves of the underlying equation (\ref{Pmz1z2}), where $(\kappa_1, \kappa_2, \kappa_3)=(1, 2, 3)$ here. It is seen that these root curves comprise two disjoint rings. Thus, they are not degenerate or empty, meeting the second condition of Theorem~\ref{Theorem1}. Then, this theorem predicts that the rational solution (\ref{DSAQ}) would exhibit spatially-bounded rogue waves. To confirm this prediction, we plot in the upper row of Fig.~\ref{f:fig4} this solution $|A(x, y, t)|$ at four time values of $-4, -2, 0$ and 4. In the $t=-4$ panel, we see four lumps on a constant background. In the $t=-2$ and 0 panels, we see that these lumps move pairwise closer, and at the same time, a spatially-bounded rogue wave in the shape of two disjoint rings appears, with each ring linking two lumps as their horizontal left and right edges. Peak amplitudes of these two rogue rings are about three times the constant background, which are reached at the same time $t=0$ for both rings. In the $t=4$ panel, these two rogue rings disappear, and the lumps move away from each other. Overall, we see that a spatially-bounded rogue wave in the shape of two disjoint rings does appear in this solution, confirming the prediction of Theorem~\ref{Theorem1}.

The reader may notice that the shape of the spatially-bounded rogue wave in the upper row of Fig.~\ref{f:fig4} closely resembles that of the root curves in Fig.~\ref{f:fig3}(a). This is certainly not an accident. As we will show in later text, the crests of this rogue wave are just a scaled version of those root curves. This phenomenon for the special case of a single large internal parameter has been reported in our earlier work \cite{YangCurve2024}.

\textbf{Example 2. } In our second example, we choose the same $N$ and $\Lambda$ values of Example 1 but different  $\textbf{\emph{a}}$ parameters. Specifically, we take
\[ \label{paraFig3b}
N=4, \quad \Lambda=(2, 3, 4, 5), \quad R=8, \quad a_2=0, \quad a_3=R^3, \quad a_4=R^4, \quad a_5=3R^5
\]
in the higher-order rational solution (\ref{DSAQ}). Notice that the only change in these parameters from Example~1 is a different $a_4$ value. Since this order-index vector $\Lambda$ is the same as that in Example~1, the first condition of Theorem~\ref{Theorem1} is then satisfied. To check the second (root curve) condition, we plot in Fig.~\ref{f:fig3}(b) the root curve of the underlying equation (\ref{Pmz1z2}), where $(\kappa_1, \kappa_2, \kappa_3)=(1, 1, 3)$ here. It is seen that this root curve has a sideways heart shape and is not degenerate or empty, thus meeting the second condition of Theorem~\ref{Theorem1}. Then, this theorem predicts that the rational solution (\ref{DSAQ}) should also admit spatially-bounded rogue waves. To confirm this prediction, we show in the lower row of Fig.~\ref{f:fig4} this solution $|A(x, y, t)|$ at four time values of $-4, -2, 0$ and 4. Graphs of this solution share many  features of those in the upper row of this figure. Particularly, a spatially-bounded rogue wave indeed appears, confirming the prediction of Theorem~\ref{Theorem1}. The main difference between graphs of these two solutions is that, in the lower row, the spatially-bounded rogue wave is in the shape of a sideways heart instead of two separate rings. In this solution, the shape of the rogue wave that appears closely resembles that of the underlying root curve in Fig.~\ref{f:fig3}(b), which is not surprising as we have mentioned in Example 1.

\begin{figure}[htb]
\begin{center}
\includegraphics[scale=0.27, bb=1440 0 300 570]{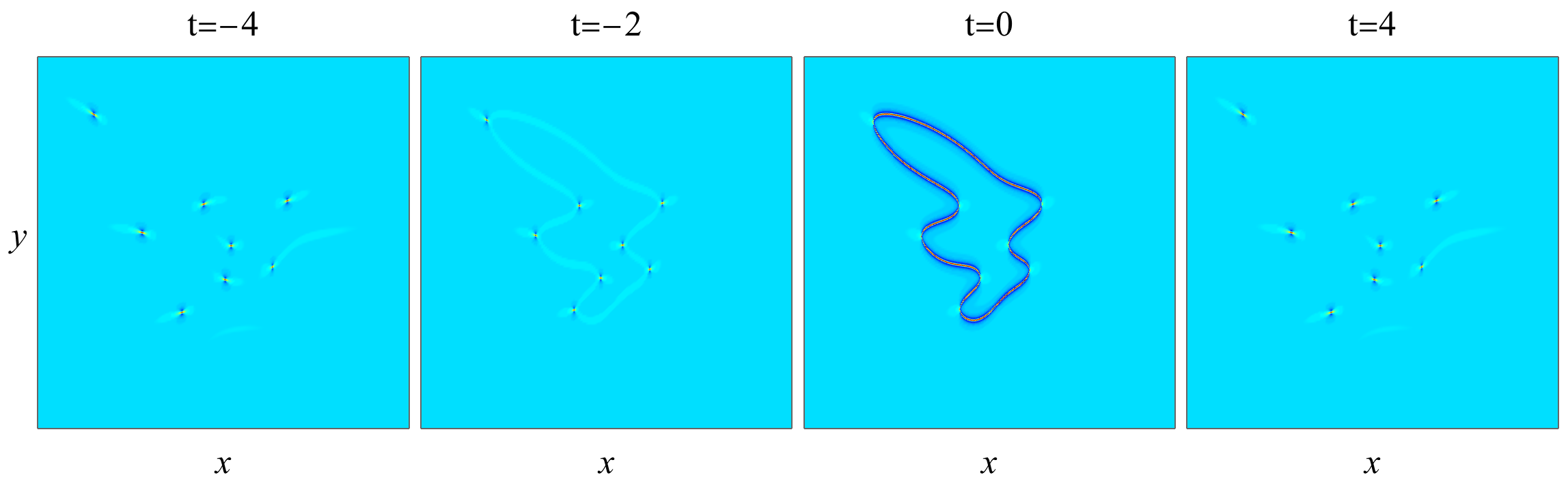}

\includegraphics[scale=0.27, bb=1440 0 300 565]{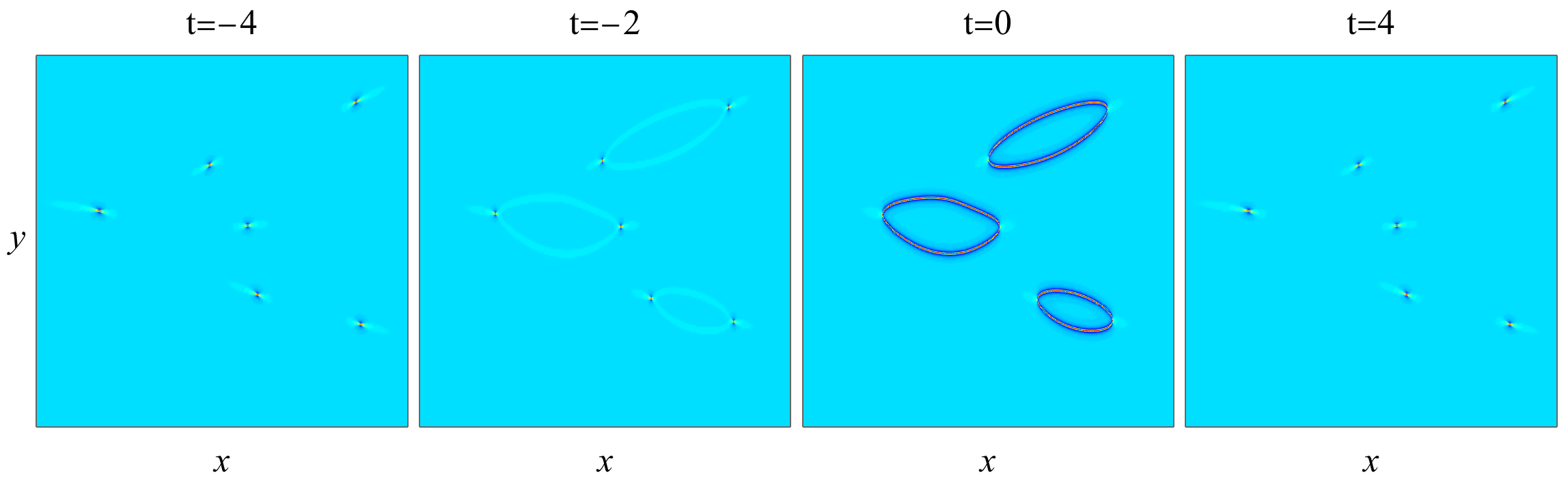}
\caption{Two more spatially-bounded rogue waves $|A(x, y, t)|$ (Examples 3 and 4) to confirm Theorem~\ref{Theorem1}. These graphs are obtained by plotting rational solutions (\ref{DSAQ}) with parameter values (\ref{paraFig4a}) (upper row) and (\ref{paraFig4b}) (lower row) at four time values of $t=-4, -2, 0$ and 4. In upper panels, $-900\le x\le 900$ and $-80\le y\le 100$; in lower panels, $-800\le x\le 800$ and $-100\le y\le 100$. } \label{f:fig5}
\end{center}
\end{figure}

\textbf{Example 3. } In our third example, we choose parameters
\[ \label{paraFig4a}
N=4, \hspace{0.15cm} \Lambda=(2, 3, 8, 9), \hspace{0.15cm} R=8, \hspace{0.15cm} a_2=0, \hspace{0.15cm}a_3=R^3, \hspace{0.15cm} a_4=R^4, \hspace{0.15cm} a_5=R^5, \hspace{0.15cm} a_6=a_7=a_8=0, \hspace{0.15cm} a_9=R^9
\]
in the higher-order rational solution (\ref{DSAQ}). This order-index vector $\Lambda$ meets the first condition of  Theorem~\ref{Theorem1}. To check its second condition, we plot in Fig.~\ref{f:fig3}(c) the root curve of the underlying equation (\ref{Pmz1z2}), where $(\kappa_1, \dots, \kappa_7)=(1, 1, 1, 0, 0, 0, 1)$ here. It is seen that this root curve forms a connected irregular shape. Thus, it is not degenerate or empty, meeting the second condition of Theorem~\ref{Theorem1}. Then, this theorem predicts that the rational solution (\ref{DSAQ}) would exhibit spatially-bounded rogue waves. To confirm this prediction, we show in the upper row of Fig.~\ref{f:fig5} this solution $|A(x, y, t)|$ at four time values of $-4, -2, 0$ and 4. In the $t=-4$ panel, we see eight lumps on a constant background. In the $t=-2$ and 0 panels, we see that a spatially-bounded rogue wave in the shape of an irregular closed curve appears that links these eight lumps together, and peak amplitudes on the crests of this irregular-shaped rogue wave are about three times the constant background. In the $t=4$ panel, this rogue wave disappears and the lumps recover themselves. Overall, we see that a spatially-bounded rogue wave in the shape of an irregular closed curve does appear in this solution, confirming the prediction of Theorem~\ref{Theorem1}. As expected, the rogue shape that arises closely resembles the underlying root curve in Fig.~\ref{f:fig3}(c).

\textbf{Example 4. } In our fourth example, we choose parameters
\[ \label{paraFig4b}
N=6,\hspace{0.15cm}  \Lambda=(2, 3, 4, 5, 6, 7), \hspace{0.15cm}  R=8, \hspace{0.15cm} a_2= a_3=0, \hspace{0.15cm} a_4=-R^4, \hspace{0.15cm} a_5=2R^5, \hspace{0.15cm} a_6=2R^6, \hspace{0.15cm} a_7=2R^7
\]
in the higher-order rational solution (\ref{DSAQ}). This order-index vector $\Lambda$ meets the first condition of  Theorem~\ref{Theorem1}. To check its second condition, we plot in Fig.~\ref{f:fig3}(d) the root curves of the underlying equation (\ref{Pmz1z2}), where $(\kappa_1, \dots, \kappa_5)=(0, -1, 2, 2, 2)$ here. It is seen that these root curves comprise three disjoint rings. Thus, they are not degenerate or empty, meeting the second condition of Theorem~\ref{Theorem1}. Then,  this theorem predicts that the rational solution (\ref{DSAQ}) should exhibit spatially-bounded rogue waves. To confirm this prediction, we show in the lower row of Fig.~\ref{f:fig5} this solution $|A(x, y, t)|$ at four time values of $-4, -2, 0$ and 4. In the $t=-4$ panel, we see six lumps on a constant background. In the $t=-2$ and 0 panels, we see that a spatially-bounded rogue wave in the shape of three disjoint rings appears, with each ring linking two of the lumps. Peak amplitudes of these three rogue rings are about three times the constant background, which are reached at the same time $t=0$ for all three rings. In the $t=4$ panel, these three rogue rings disappear and the lumps recover themselves. Overall, a spatially-bounded rogue wave in the shape of three disjoint rings appears in this solution, confirming the prediction of Theorem~\ref{Theorem1}. Again, the rogue shape that appears closely resembles the underlying root curves in Fig.~\ref{f:fig3}(d) as expected.

If the two conditions of Theorem~\ref{Theorem1} are not met, what will happen to the higher-order rational solution (\ref{DSAQ})? We have already seen that in the line rogue wave (\ref{1strogue}) and the second-order rational solution (\ref{2ndrogue}), whose order-index vectors do not meet the first condition of Theorem~\ref{Theorem1}, spatially-unbounded rogue waves arise (see Fig.~\ref{f:fig1}). Below, we will examine two more examples.

\begin{figure}[htb]
\begin{center}
\includegraphics[scale=0.27, bb=1440 0 300 570]{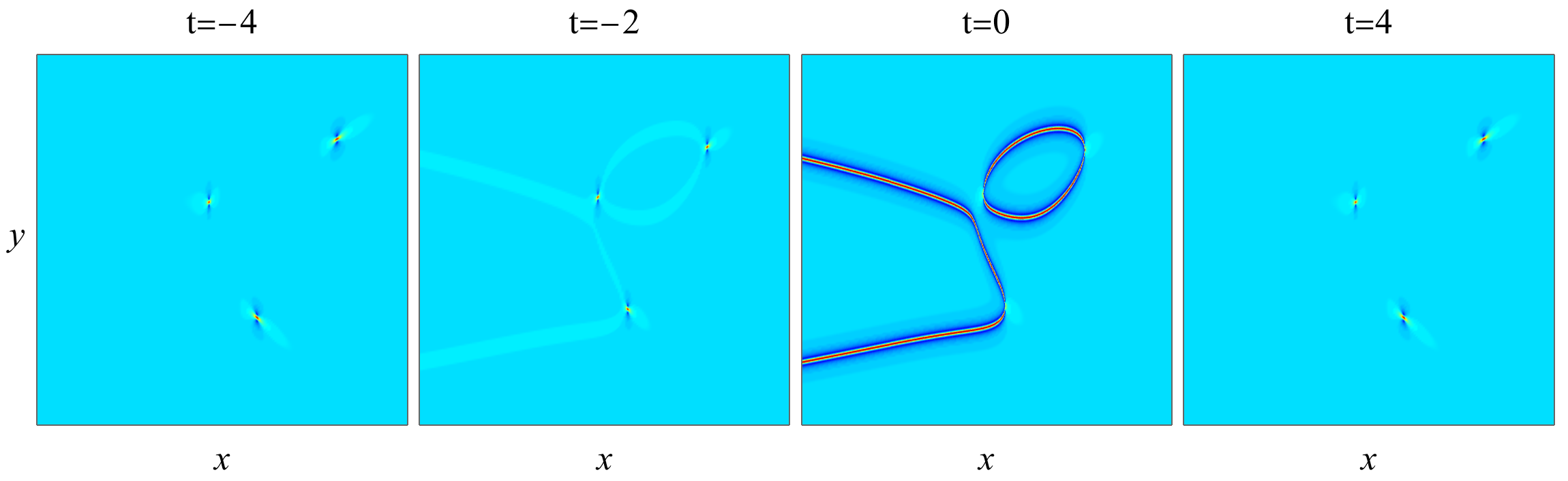}

\includegraphics[scale=0.27, bb=1440 0 300 570]{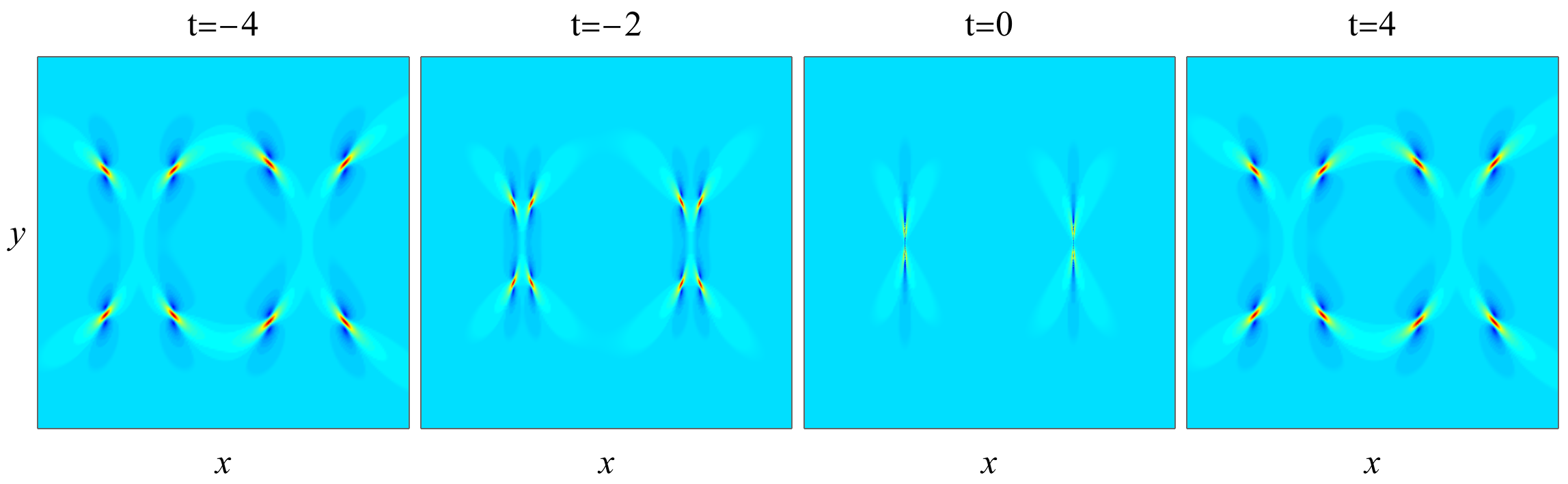}
\caption{Two higher-order rational solutions $|A(x, y, t)|$ (Examples 5 and 6) which do not meet the conditions of Theorem~\ref{Theorem1}. These graphs are obtained by plotting rational solutions (\ref{DSAQ}) with parameter values (\ref{paraFig5a}) (upper row) and (\ref{paraFig5b}) (lower row) at four time values of $t=-4, -2, 0$ and 4. In upper panels, $-800\le x\le 600$ and $-42\le y\le 50$; in lower panels, $-400\le x\le 400$ and $-20\le y\le 20$. } \label{f:fig6}
\end{center}
\end{figure}

\textbf{Example 5. } In our fifth example, we choose parameters
\[ \label{paraFig5a}
N=3, \hspace{0.15cm} \Lambda=(2, 3, 4), \hspace{0.15cm} R=8, \hspace{0.15cm} a_2=0, \hspace{0.15cm}a_3=R^3, \hspace{0.15cm} a_4=-R^4
\]
in the higher-order rational solution (\ref{DSAQ}). This order-index vector $\Lambda$ does not meet the first condition of Theorem~\ref{Theorem1}, but its second condition is satisfied which we have checked. To learn what happens in this rational solution, we show in the upper row of Fig.~\ref{f:fig6} this solution $|A(x, y, t)|$ at four time values of $-4, -2, 0$ and 4. In the $t=-4$ panel, we see three lumps on a constant background. In the $t=-2$ and 0 panels, we see a rogue wave appearing. This rogue wave comprises two pieces: one is a spatially-bounded ring at the upper-right corner, and the other has a sideways U shape on the left side that is spatially-unbounded. At large negative $x$ values, this sideways U piece approaches a left-facing parabola. In the $t=4$ panel, both pieces of this rogue wave disappear and the three lumps recover themselves. In short, this higher-order rational solution (\ref{DSAQ}) exhibits a rogue wave that is spatially-unbounded overall, even though it does contain a spatially-bounded rogue piece. The crests of this rogue wave, including both spatially-bounded and spatially-unbounded components, closely resemble the shapes of root curves of the underlying equation (\ref{Pmz1z2}), similar to the previous four examples.

As Example 5 shows, in general if the second condition of Theorem~\ref{Theorem1} is met but not the first, then rogue waves would still arise but are spatially-unbounded as a whole, and their crests can be predicted by root curves of the underlying equation (\ref{Pmz1z2}).

\textbf{Example 6. } In our last example, we choose parameters
\[ \label{paraFig5b}
N=2, \hspace{0.15cm} \Lambda=(4,5), \hspace{0.15cm} R=8, \hspace{0.15cm} a_2=a_3=0, \hspace{0.15cm} a_4=-R^4, \hspace{0.15cm} a_5=0
\]
in the higher-order rational solution (\ref{DSAQ}). In this case, $(\kappa_1, \kappa_2, \kappa_3)=(0, -1, 0)$. For these $\Lambda$ and $\kappa_j$ values, the corresponding $\mathcal{P}_{\Lambda}(z_1, z_2)$ polynomial is given in Eq.~(\ref{Pdege}), whose root curves are degenerate. Thus, the first condition of Theorem~\ref{Theorem1} is met but not the second. In this case, to learn what happens in this rational solution, we show in the lower row of Fig.~\ref{f:fig6} this solution $|A(x, y, t)|$ at four time values of $-4, -2, 0$ and 4. In the $t=-4$ panel, we see eight lumps on a constant background. As time increases, the left four lumps move closer together, and the right four lumps move close together. Simultaneously, all lumps shrink in size.  At $t=0$, the left four lumps coalesce and the right four lumps also coalesce. Afterwards, the process is reversed, and the eight lumps recover themselves (see the $t=4$ panel). Overall, no rogue waves appear. This phenomenon is also true in general. That is, if the second condition of Theorem~\ref{Theorem1} is not met, then the higher-order rational solution (\ref{DSAQ}) does not exhibit rogue waves, spatially-bounded or not.

\subsection{Proof of Theorem~\ref{Theorem1}} \label{sec:proof}
Now, we prove Theorem~\ref{Theorem1}. This proof starts with the asymptotic analysis of the higher-order rational solution (\ref{DSAQ}) under multiple large internal parameters (\ref{acond}) when $a_2=O(1)$ and $t=O(1)$.

First, we notice from Eq.~(\ref{xrijplus2}) that
\[
x_1^+(k)=y- 2 \textrm{i} t + k, \quad x_2^+=\frac{1}{2}x +a_2,  \quad x_3^+=\frac{1}{6}y -\frac{4}{3} \textrm{i} t+a_3, \quad x_4^+=\frac{1}{24}x+a_4,
\]
and so on. According to our parameter conditions, $a_2=O(1)$, and the other internal parameters $a_j=\kappa_{j-2}R^j$ with $R\gg 1$. In this case,  let $x=O(R^2)$, $y=O(R)$, $t=O(1)$, and denote
\[ \label{xyz1z2b}
x = 2 z_1 R^2, \quad y = z_2 R,
\]
where $z_1$ and $z_2$ are real and $O(1)$. Then,
\begin{eqnarray*}
&& S_n\left[\textbf{\emph{x}}^{+}(k)+\nu \textbf{\emph{s}}\right]=S_n ( y-2\textrm{i}t+k,  \frac{1}{2}x+a_2+\nu s_2, \hspace{0.06cm} \frac{1}{6}y -\frac{4}{3} \textrm{i} t+a_3, \hspace{0.06cm}  \frac{1}{24}x+a_4+\nu s_4, \cdots )  \\
&& \sim S_n(y, \frac{1}{2}x, a_3, a_4, \cdots) =S_n(z_2R, z_1R^2, \kappa_1R^3, \kappa_2R^4, \cdots) = R^{n}  S_n\left(z_2, z_1, \kappa_1, \kappa_2, \cdots  \right)=R^{n} H_n(z_1, z_2).
\end{eqnarray*}
Thus,
\[ \label{SnplusRR}
S_n\left[\textbf{\emph{x}}^{+}(k)+\nu \textbf{\emph{s}}\right]\sim R^n H_n(z_1, z_2), \quad R\gg 1.
\]
Similarly, we can also show that
\[ \label{SnminusRR}
S_n\left[\textbf{\emph{x}}^{-}(k)+\nu \textbf{\emph{s}}\right]\sim R^n H_n(z_1, z_2), \quad R\gg 1.
\]

Next, we use a technique of \cite{OhtaJY2012,YangCurve2024} to rewrite the determinant $\tau_k$ in Eq.~(\ref{deftaunk}) as a Laplace expansion
\begin{eqnarray} \label{sigmanLap}
&& \hspace{-0.8cm} \tau_k=\sum_{0\leq\nu_{1} < \nu_{2} < \cdots < \nu_{N}\leq n_N}
\det_{1 \leq i, j\leq N} \left(\frac{1}{2^{\nu_j}} S_{n_i-\nu_j}({\textbf{\emph{x}}}^{+}(k)+\nu_j \textbf{\emph{s}}) \right)  \times \det_{1 \leq i, j\leq N}\left(\frac{1}{2^{\nu_j}}S_{n_i-\nu_j} ({\textbf{\emph{x}}}^{-}(k)+ \nu_j \textbf{\emph{s}} )\right).
\end{eqnarray}
The highest-power term of $R$ in $\tau_k$ comes from the index choices of $\nu_{j}=j-1$. Then, using Eqs.~(\ref{SnplusRR})-(\ref{SnminusRR}), we can show that the highest $R$-power term of $\tau_k$ is
\begin{equation} \label{taukmax}
\tau_k \sim 2^{-N(N-1)} \hspace{0.05cm} R^{2\beta} \left[\mathcal{P}_{\Lambda}(z_1, z_2)\right]^2,
\quad \quad R \gg 1,
\end{equation}
where $\mathcal{P}_{\Lambda}(z_1, z_2)$ is the double-real-variable polynomial defined in Eq.~(\ref{DoubleRealPolydef}), and
\[  \label{defbeta}
\beta=n_1+n_2+\dots+n_N-\frac{1}{2}N(N-1).
\]
Inserting this leading-order term of $\tau_k$ into the solution $A(x,y,t)=\sqrt{2}\tau_1/\tau_0$, we see that this $A$ solution is approximately $\sqrt{2}$, i.e., it is on the constant background, except at or near $(x, y)$ locations where $\mathcal{P}_{\Lambda}(z_1, z_2)=0$, or equivalently,
\[   \label{ycdef2}
\mathcal{P}_{\Lambda}\left(\frac{x}{2R^{2}}, \frac{y}{R}\right)=0
\]
in view of the connection (\ref{xyz1z2b}) between $(z_1, z_2)$ and $(x, y)$. We call the $(x, y)$ locations where Eq.~(\ref{ycdef2}) holds as the critical curve. Apparently, this critical curve is just a stretching of the root curve of (\ref{Pmz1z2}) along the $z_1$ and $z_2$ directions. According to the second condition of Theorem~\ref{Theorem1}, this critical curve is not degenerate or empty, i.e., it is indeed a curve, or it contains a curve if it contains isolated points as well. If $(x, y)$ is not in the $O(1)$ neighborhood of this critical curve (including isolated points), the solution $A(x,y,t)$ approaches the constant background $\sqrt{2}$ for large $R$. Thus, nontrivial dynamics of the solution can only occur in the $O(1)$ neighborhood of this critical curve.

We will show later in Sec.~\ref{sec:curve} that in the $O(1)$ neighborhood of this critical curve (meaning its curvy part, not isolated points if they also exist), a rogue wave indeed arises. Due to the second condition of Theorem~\ref{Theorem1}, this curvy component of the critical curve exists, which guarantees the appearance of a rogue wave. A result similar to this has been reported in our earlier work \cite{YangCurve2024} for the special case of a single large internal parameter. The key question now, which was not considered in \cite{YangCurve2024} and is the focus of this paper, is under what conditions this rogue wave is spatially bounded. We will show in the rest of this section that the first condition of Theorem~\ref{Theorem1} guarantees the spatial boundedness of this rogue wave.

Since this rogue wave lies in the $O(1)$ neighborhood of the critical curve (\ref{ycdef2}), whether this rogue wave is spatially bounded naturally depends on whether this critical curve is bounded. Since this critical curve is a simple stretching of the root curve of Eq.~(\ref{Pmz1z2}), then whether this rogue wave is spatially bounded depends on whether the root curve of (\ref{Pmz1z2}) is bounded. Next, we determine under what conditions the root curve of (\ref{Pmz1z2}) is bounded.

\subsubsection{Conditions for bounded root curves}

To determine under what conditions the root curve of Eq.~(\ref{Pmz1z2}) is bounded, we examine when this root curve is unbounded. When it is unbounded, it is easy to see that the $z_1$ value of this curve must be unbounded. The reason is that, if $z_1$ is bounded, then for the root curve to be unbounded, $z_2$ would have to be unbounded. But since the highest-power term in the Schur polynomial $H_n(z_1, z_2)$ is $z_2^n/n!$, which can be readily seen from Eq.~(\ref{AMthetak}), the highest-power term in the double-real-variable polynomial $\mathcal{P}_{\Lambda}(z_1, z_2)$ of Eq.~(\ref{DoubleRealPolydef}) then is a constant multiplying $z_2^{\beta}$, where $\beta$ is as given in Eq.~(\ref{defbeta}). If $z_1$ is bounded but $z_2$ unbounded, the equation $\mathcal{P}_{\Lambda}(z_1, z_2)=0$ cannot be satisfied because there is nothing to balance its highest-power term $z_2^{\beta}$. Thus, if the root curve is unbounded, then the $z_1$ value of this root curve must be unbounded.

Now, we determine under what conditions the $z_1$ value of the root curve is unbounded. We first consider the case of $z_1$ unbounded along its positive direction, i.e., $z_1\to +\infty$ on this root curve. For large positive $z_1$, we rescale $\hat{\epsilon}=\sqrt{z_1}\epsilon$ in Eq.~(\ref{AMthetak}) and rewrite that equation as
\begin{equation}\label{AMthetak2}
\sum_{n=0}^{\infty} \widehat{H}_n(\hat{z}) \hspace{0.04cm} \hat{\epsilon}^k =\exp\left( \hat{z} \hat{\epsilon} + \hat{\epsilon}^{2} +
\sum_{j=1}^{\infty} \hat{\kappa}_{j} \hat{\epsilon}^{j+2}\right),
\end{equation}
where
\[
\widehat{H}_n(\hat{z})=\frac{H_n(z_1, z_2)}{z_1^{n/2}}, \quad
\hat{z}=\frac{z_2}{\sqrt{z_1}}, \quad \hat{\kappa}_j=\frac{\kappa_j}{z_1^{(j+2)/2}}.
\]
Under this rescaling, the double-real-variable polynomial $\mathcal{P}_{\Lambda}(z_1, z_2)$ in Eq.~(\ref{DoubleRealPolydef}) becomes
\[\label{DoubleRealPolydef2}
\mathcal{P}_{\Lambda}(z_1, z_2)=z_1^{\beta/2} \hspace{0.06cm} \widehat{\mathcal{P}}_{\Lambda}(\hat{z}),
\]
where $\beta$ is as given in Eq.~(\ref{defbeta}),
\begin{equation} \label{DoubleRealPolydef3}
\widehat{\mathcal{P}}_{\Lambda}(\hat{z}) =\det_{1\le i\le N}\left[
         \widehat{H}_{n_i}(\hat{z}), \hspace{0.05cm} \widehat{H}_{n_i-1}(\hat{z}), \hspace{0.05cm}  \cdots, \hspace{0.05cm}  \widehat{H}_{n_i-N+1}(\hat{z}) \right],
\end{equation}
and $ \widehat{H}_{n}(\hat{z})\equiv 0$ if $n<0$. When $z_1\to +\infty$, all $\hat{\kappa}_j\to 0$. In this limit, Eq.~(\ref{AMthetak2}) becomes
\begin{equation}\label{AMthetak3}
\sum_{n=0}^{\infty} \widetilde{H}_n(\hat{z}) \hat{\epsilon}^k =\exp\left( \hat{z} \hat{\epsilon} + \hat{\epsilon}^{2} \right),
\end{equation}
and Eq.~(\ref{DoubleRealPolydef2}) becomes
\[\label{DoubleRealPolydef2b}
\mathcal{P}_{\Lambda}(z_1, z_2)=z_1^{\beta/2} \hspace{0.06cm} \widetilde{\mathcal{P}}_{\Lambda}(\hat{z}),
\]
where
\begin{equation} \label{DoubleRealPolydef3b}
\widetilde{\mathcal{P}}_{\Lambda}(\hat{z}) =\det_{1\le i\le N}\left[
         \widetilde{H}_{n_i}(\hat{z}), \hspace{0.05cm} \widetilde{H}_{n_i-1}(\hat{z}), \hspace{0.05cm}  \cdots, \hspace{0.05cm}  \widetilde{H}_{n_i-N+1}(\hat{z}) \right].
\end{equation}
Eq.~(\ref{AMthetak3}) shows that $\widetilde{H}_n(\hat{z})$ is just the familiar Hermite polynomial under minor notational differences. It is easy to see from this equation that $\widetilde{H}'_n(\hat{z})= \widetilde{H}_{n-1}(\hat{z})$, where the prime denotes differentiation. Thus, $\widetilde{\mathcal{P}}_{\Lambda}(\hat{z})$ in Eq.~(\ref{DoubleRealPolydef3b}) is just the Wronskian-Hermite polynomial.

Eq.~(\ref{DoubleRealPolydef2b}) shows that, in order for the root curve of the equation $\mathcal{P}_{\Lambda}(z_1, z_2)=0$ to be unbounded along the positive $z_1$ direction, the Wronskian-Hermite polynomial  $\widetilde{\mathcal{P}}_{\Lambda}(\hat{z})$ must admit real roots $\hat{z}_0$, in which case $z_2/\sqrt{z_1}\to \hat{z}_0$ on that unbounded root curve branch.

Using similar arguments, we find that in order for the root curve of the equation $\mathcal{P}_{\Lambda}(z_1, z_2)=0$ to be unbounded along the negative $z_1$ direction, the Wronskian-Hermite polynomial  $\widetilde{\mathcal{P}}_{\Lambda}(\hat{z})$ must admit purely imaginary roots $\hat{z}_0$, in which case $z_2/\sqrt{z_1}\to \hat{z}_0$ on that unbounded root curve branch.

Combining these results, we see that the root curve of the equation $\mathcal{P}_{\Lambda}(z_1, z_2)=0$ is unbounded if and only if the Wronskian-Hermite polynomial  $\widetilde{\mathcal{P}}_{\Lambda}(\hat{z})$ admits real or purely-imaginary roots. Then, this root curve is bounded if and only if the Wronskian-Hermite polynomial  $\widetilde{\mathcal{P}}_{\Lambda}(\hat{z})$ does not admit any real or purely-imaginary roots.

\subsubsection{Conditions for nonexistence of real or purely-imaginary roots in Wronskian-Hermite polynomials}
The above analysis converted the question of spatially-bounded rogue waves to the following question: under what conditions does the Wronskian-Hermite polynomial  $\widetilde{\mathcal{P}}_{\Lambda}(\hat{z})$ not admit any real and purely-imaginary roots?

To answer this question, the result of Adler in \cite{Adler1994} will play a key role. In this paper, Adler gave the necessary and sufficient conditions for the Wronskian-Hermite polynomial $\widetilde{\mathcal{P}}_{\Lambda}(\hat{z})$ to have no purely-imaginary roots. It is noted that the Hermite polynomial in Adler's work used the conventional generating function different from ours in (\ref{AMthetak3}) by a negative sign in front of $\hat{\epsilon}^2$ of the exponent. As such, purely-imaginary roots in our Wronskian-Hermite polynomial $\widetilde{\mathcal{P}}_{\Lambda}(\hat{z})$ are real roots in Adler's Wronskian-Hermite polynomial. Translating Adler's results to our notations, his results say that the Wronskian-Hermite polynomial $\widetilde{\mathcal{P}}_{\Lambda}(\hat{z})$ in Eq.~(\ref{DoubleRealPolydef3b}) has no purely-imaginary roots (including zero) if and only if the order-index vector $\Lambda=(n_1, n_2, \dots, n_N)$ with positive integer elements is built by concatenation of segments of consecutive integers of even length, such as $(3, 4)$, $(2, 3, 4, 5)$, $(2, 3, 7, 8, 9, 10)$, and so on. This means that $N$ has to be even, and the order-index vector $\Lambda$ can be written as
\[ \label{LambdaAdler}
\Lambda=(j_1, j_1+1, j_2, j_2+1, \cdots, j_m, j_m+1),
\]
where $m=N/2$, $j_1, \cdots, j_m$ are positive integers, and $j_k+2\le j_{k+1}$ for each $1\le k\le m-1$.

Now, we also need to find conditions for the Wronskian-Hermite polynomial $\widetilde{\mathcal{P}}_{\Lambda}(\hat{z})$ to have no real roots. This can be done by using the theory of symmetric functions \cite{Murnaghan1937,Chakravarty2022KPI}. For this purpose, it would be necessary to introduce the Young diagram $Y = (i_1, i_2, \dots, i_N)$, or a partition, of length $N$, such that $i_1 \ge i_2 \ge \dots \ge i_N > 0$. The Schur function $W_Y(\emph{\textbf{x}})$, for vector $\emph{\textbf{x}}=(x_1, x_2, \dots)$ and Young diagram $Y = (i_1, i_2, \dots, i_N)$, is defined by
\[
W_Y(\emph{\textbf{x}})=\det_{1\le j, k\le N} [S_{i_j -j+k}(\emph{\textbf{x}})],
\]
where elementary Schur polynomials $S_j(\emph{\textbf{x}})$ are as defined in Eq.~(\ref{Elemgenefunc}). In terms of this Schur function, the Wronskian-Hermite polynomial $\widetilde{\mathcal{P}}_{\Lambda}(\hat{z})$ as defined in Eq.~(\ref{DoubleRealPolydef3b}) can be written as
\[ \label{PWY}
\widetilde{\mathcal{P}}_{\Lambda}(\hat{z})=W_Y(\emph{\textbf{x}}),
\]
where $\emph{\textbf{x}}=(\hat{z}, 1, 0, \dots)$, and the Young diagram $Y = (i_1, i_2, \dots, i_N)$ of the order-index vector $\Lambda=(n_1, n_2, \dots, n_N)$ is given by
\[\label{Indexrelation}
i_{N-(j-1)}=n_j-(j-1), \quad j=1, \dots, N.
\]
For example, when $\Lambda=(4, 5)$, $Y=(4,4)$; and when $\Lambda=(2, 3, 4, 5)$, $Y=(2,2,2,2)$.

The Young diagram $Y = (i_1, i_2, \dots, i_N)$ is often displayed as a rectangular array of left-justified boxes such that the $k$-th row from the top contains $i_k$ boxes, $k = 1,...,N$. Thus, the Young diagram consists of $N$ rows and a total number of $|Y|$ boxes, where $|Y|\equiv i_1+\cdots+i_N$. The conjugate $Y'$ of a partition $Y$ is a partition whose Young diagram is the transpose of the original one obtained by interchanging its rows and columns. Borrowing this notation, we define the conjugate $\Lambda'$ of the order-index vector $\Lambda$ as one whose Young diagram is the conjugate of $\Lambda$'s Young diagram. For example, when $\Lambda=(4, 5)$, $Y=(4,4)$. Thus, $Y'=(2,2,2,2)$ and $\Lambda'=(2,3,4,5)$.

A well known result from the theory of symmetric functions \cite{Murnaghan1937,Chakravarty2022KPI} is  the following \emph{involution symmetry} among Schur functions of a given partition $Y$ and its conjugate $Y'$:
\[\label{involution}
W_{Y'}(\emph{\textbf{x}})= W_Y(\omega(\emph{\textbf{x}})),\quad  \omega(x_j)= (-1)^{j-1}x_{j}.
\]
For the vector $\emph{\textbf{x}}=(\hat{z}, 1, 0, \dots)$ in Eq.~(\ref{PWY}), this involution symmetry yields
\[\label{involution2}
W_{Y'}(\emph{\textbf{x}})= W_Y(\tilde{\emph{\textbf{x}}}),
\]
where $\tilde{\emph{\textbf{x}}}=(\hat{z}, -1, 0, \dots)$. It is easy to see from the definition of elementary Schur polynomials that $S_n(\tilde{\emph{\textbf{x}}})=(-{\rm{i}})^n S_n(\hat{\emph{\textbf{x}}})$, where $\hat{\emph{\textbf{x}}}=({\rm{i}}\hat{z}, 1, 0, \dots)$. Thus, the involution symmetry (\ref{involution2}) leads to
$W_{Y'}(\emph{\textbf{x}})= (-{\rm{i}})^{|Y|} W_Y(\hat{\emph{\textbf{x}}})$, which means that
\[ \label{WHinvolution}
\widetilde{\mathcal{P}}_{\Lambda'}(\hat{z})=(-{\rm{i}})^{|Y|}\hspace{0.04cm} \widetilde{\mathcal{P}}_{\Lambda}({\rm{i}}\hat{z}).
\]
Note that $|Y|$ here is equal to $\beta$ as defined in Eq.~(\ref{defbeta}) in view of the connection (\ref{Indexrelation}). Because of the above relation (\ref{WHinvolution}), real roots of $\widetilde{\mathcal{P}}_{\Lambda}(\hat{z})$ are purely-imaginary roots of $\widetilde{\mathcal{P}}_{\Lambda'}(\hat{z})$. Thus,  $\widetilde{\mathcal{P}}_{\Lambda}(\hat{z})$ having no real roots is equivalent to the Wronskian-Hermit polynomial $\widetilde{\mathcal{P}}_{\Lambda'}(\hat{z})$ for the conjugate order index $\Lambda'$ having no purely-imaginary roots.

So, we need to determine under what conditions the Wronskian-Hermit polynomial $\widetilde{\mathcal{P}}_{\Lambda'}(\hat{z})$ has no purely-imaginary roots. For the order-index vector $\Lambda$ in Eq.~(\ref{LambdaAdler}), its Young diagram is
\[ \label{Ymu}
Y=(i_1,i_1, i_2, i_2, \dots, i_m, i_m),
\]
where $i_k=j_{m+1-k}-2(m-k)$ for each $1\le k\le m$. Notice that $i_1\ge i_2\ge \cdots \ge i_m$. This form of the partition is called a double partition in the literature, where each element appears twice, or even number of times if some $i_k$'s are the same. Since Adler's result in \cite{Adler1994} shows that (\ref{LambdaAdler}) is the necessary and sufficient order-index condition for the Wronskian-Hermite polynomial $\widetilde{\mathcal{P}}_{\Lambda}(\hat{z})$ to have no purely-imaginary roots, this means that for $\widetilde{\mathcal{P}}_{\Lambda}(\hat{z})$ to have no purely-imaginary roots, the necessary and sufficient condition is that the Young diagram associated with $\Lambda$ is a double partition.

The conjugate of the partition (\ref{Ymu}) is
\[ \label{Y'mu}
Y'=[(2m)^{i_m}, (2m-2)^{i_{m-1}-i_m}, \cdots, 6^{i_3-i_4}, 4^{i_2-i_3}, 2^{i_1-i_2}],
\]
where the power of a number in this $Y'$ means that this number repeats that power times (for instance, $2m$ repeats $i_m$ times in this $Y'$). Then, utilizing Adler's results quoted above, the necessary and sufficient condition for $\widetilde{\mathcal{P}}_{\Lambda'}(\hat{z})$ to have no purely-imaginary roots is that the Young diagram $Y'$ in (\ref{Y'mu}) is a double partition. For this to be true, the powers in $Y'$ of (\ref{Y'mu})  must be all even integers, which means that all $(i_1, i_2, \cdots, i_m)$ must be even integers. Then, in view of the connection between $i_k$ and $j_k$ below Eq.~(\ref{Ymu}), we see that all $(j_1, j_2, \cdots, j_m)$ must be even integers as well. Putting all these results together, we conclude that the Wronskian-Hermite polynomial  $\widetilde{\mathcal{P}}_{\Lambda}(\hat{z})$ does not admit real or purely-imaginary roots if and only if its order-index vector $\Lambda$ is of the form (\ref{LambdaAdler}), where every $j_k$ is an even integer. In other words, the order-index vector $\Lambda$ is a concatenation of pairs of the form $(2n, 2n+1)$, where $n$ is a positive integer. This completes the proof of Theorem~\ref{Theorem1}. $\Box$

We note by passing that under mild conditions, the number of simple purely-imaginary roots in the Wronskian-Hermite polynomial  $\widetilde{\mathcal{P}}_{\Lambda}(\hat{z})$ with an arbitrary order-index vector $\Lambda$ can be explicitly calculated \cite{GarciaFerrero2015}. Utilizing the involution property (\ref{WHinvolution}), the number of simple real roots in these polynomials can be explicitly calculated as well.

\section{Asymptotic reductions of rogue waves}

The original formulae (\ref{DSAQ})-(\ref{schurcoeffsr}) for DSI's higher-order rational solutions are quite complicated, and it is hard to see their dynamical features. In addition, it is hard to see the shape and location of rogue waves from those formulae if such waves do arise. Even with their solution graphs such as Figs.~\ref{f:fig4}-\ref{f:fig5} at hand so that we can see the appearance of a rogue wave, it is still unclear from these graphs what happens at horizontal local edges of these rogue waves. At those local edges (for example, the two solutions in Fig.~\ref{f:fig4} contain two left edges and two right edges each), we seem to see some kind of lumps there, but the nature of those lumps is unclear yet. Is this lump similar to that in the second-order rational solution shown in Fig.~\ref{f:fig2} that deforms over time, or it is similar to the shape-preserving lump moving on a constant background that the DSI equation also admits \cite{Satsuma_Ablowitz,OhtaYangDSI}?

To clarify these questions, the asymptotic reduction of these rational solutions is in order. For single large internal parameters in these rational solutions, their asymptotic reductions were considered in \cite{YangCurve2024}. It was shown that a rogue wave with a vertical Peregrine profile appears near a critical curve, except for certain special locations of the critical curve such as its horizontal local edges. Solution behaviors near those special locations were not clarified in \cite{YangCurve2024}.

In this section, we will greatly generalize the results of \cite{YangCurve2024} and do much more. For multiple large internal parameters (\ref{acond}), we will first derive spatially uniformly-valid asymptotic expressions of these rational solutions (\ref{DSAQ}). Then, we will further simplify these asymptotic expressions near the critical curve (\ref{ycdef2}) where nontrivial dynamics occurs. Away from horizontal local edges of the critical curve, we will show that the higher-order rational solution (\ref{DSAQ}) asymptotically reduces to a vertical Peregrine rogue profile (\ref{1strogue}), with its center lying on the critical curve. More importantly, we will show that near a horizontal edge of the critical curve, the rational solution (\ref{DSAQ}) asymptotically reduces to a second-order rational solution (\ref{2ndroguetauk}) or its horizontal reflection (\ref{2ndroguetauk2}), which contains a parabola-shaped rogue component as well as a time-varying lump. This latter result clarifies the nature and dynamics of the rational solution (\ref{DSAQ}) near horizontal edges of its rogue wave, particularly regarding the lumps that appear there (see Figs.~\ref{f:fig4}-\ref{f:fig5}).

\subsection{A spatially uniformly valid asymptotic approximation}
Now, we derive asymptotic reductions of the higher-order rational solution (\ref{DSAQ}) under multiple large internal parameters (\ref{acond}) when $a_2=O(1)$ and $t=O(1)$. This calculation is a refined version of the preliminary calculations presented in Sec.~\ref{sec:proof}.  Using notations introduced there, we see that
 \begin{eqnarray}
&& S_n\left[\textbf{\emph{x}}^{+}(k)+\nu \textbf{\emph{s}}\right]=S_n ( y-2\textrm{i}t+k,  \frac{1}{2}x+a_2+\nu s_2, \hspace{0.06cm} \frac{1}{6}y -\frac{4}{3} \textrm{i} t+a_3, \hspace{0.06cm}  \frac{1}{24}x+a_4+\nu s_4, \cdots )  \nonumber \\
&& \hspace{0.02cm}  =S_n ( z_2 R-2\textrm{i}t+k, \hspace{0.05cm} z_1 R^2 +a_2+\nu s_2, \hspace{0.06cm} \frac{1}{6}z_2 R -\frac{4}{3} \textrm{i} t+\kappa_1 R^3, \hspace{0.06cm}  \frac{1}{12}z_1 R^2+\kappa_2 R^4+\nu s_4, \cdots )  \nonumber \\
&&  \hspace{0.02cm}  = R^n S_n [ z_2 +(k-2\textrm{i}t)R^{-1}, \hspace{0.05cm} z_1+(a_2+\nu s_2)R^{-2}, \hspace{0.06cm} \kappa_1+\frac{1}{6}z_2 R^{-2} -\frac{4}{3} \textrm{i} t R^{-3}, \hspace{0.06cm}  \kappa_2+\frac{1}{12}z_1 R^{-2}+\nu s_4 R^{-4}, \cdots].     \label{Snasym}
\end{eqnarray}
A similar expression can be written for $S_n\left[\textbf{\emph{x}}^{-}(k)+\nu \textbf{\emph{s}}\right]$. In view of these expressions, the main contributions to $\tau_k$'s Laplace expansion (\ref{sigmanLap}) come from four index choices of $(\nu_1, \nu_2, \dots, \nu_N)$ in that Laplace expansion that will be calculated separately below. The reason we need to account for these many contributions is that we need to calculate terms of more orders since the leading-order terms could vanish at certain spatial locations. This keeping of more-order terms is important for deriving a spatially uniformly-valid asymptotic approximation.

Next, we calculate these four main contributions to $\tau_k$'s Laplace expansion (\ref{sigmanLap}).

\vspace{0.1cm}
(1) Contributions from the index choice of $\boldsymbol{\nu}=(0, 1, \dots, N-2, N-1)$

For this index choice, utilizing the above equation (\ref{Snasym}), we see that the $S_{n_i-\nu_j} ({\textbf{\emph{x}}}^{+}(k)+ \nu_j \textbf{\emph{s}})$ determinant in the Laplace expansion (\ref{sigmanLap}) is
\begin{eqnarray}
&& \hspace{-1.25cm} \det_{1 \leq i, j\leq N} \left(\frac{1}{2^{\nu_j}} S_{n_i-\nu_j}({\textbf{\emph{x}}}^{+}(k)+\nu_j \textbf{\emph{s}}) \right) =2^{-N(N-1)/2}R^\beta \left\{ \widetilde{P}_\Lambda \left[z_1+a_2 R^{-2}, \hspace{0.06cm} z_2 +(k-2\textrm{i}t)R^{-1}, \hspace{0.05cm}  \kappa_1+\frac{1}{6}z_2 R^{-2}, \hspace{0.06cm}  \kappa_2+\frac{1}{12}z_1 R^{-2}, \cdots\right] \right.  \nonumber \\
&& \hspace{6.6cm}  \left. +s_2 B_1(z_1, z_2)  R^{-2} +O(R^{-3})\right\},    \label{contri1}
\end{eqnarray}
where the constant $\beta$ is as given in Eq.~(\ref{defbeta}), $\widetilde{P}_\Lambda(z_1, z_2, \kappa_1, \kappa_2, \dots)$ is the same function $P_\Lambda(z_1, z_2)$ as defined in Eqs.~(\ref{DoubleRealPolydef})-(\ref{AMthetak}) except to treat $\kappa_1, \kappa_2, \dots$ as variables as well, and $B_1(z_1, z_2)$ is given by
\[ \label{B1}
B_1(z_1, z_2)=\sum_{j=1}^{N-1}  j \det_{1\le i\le N}
         (H_{n_i}, H_{n_i-1}, \cdots,  \partial H_{n_i-j}/\partial z_1,  \cdots,   H_{n_i-N+1}),
\]
with $H_n=H_n(z_1, z_2)$ as defined in Eq.~(\ref{AMthetak}). The determinant in this $B_1$ formula is the determinant of Eq.~(\ref{DoubleRealPolydef}) but with its $(j+1)$-th column differentiated with respect to $z_1$. The $s_2B_1$ term in the above equation (\ref{contri1}) comes from the $\nu s_2$ term in (\ref{Snasym}) when $\nu_j$ takes on values of $1, 2, \dots, N-1$ in the left-hand-side determinant of Eq.~(\ref{contri1}).  Notice from $H_n$'s definition (\ref{AMthetak}) that $\partial H_n/\partial z_1=H_{n-2}$. This relation can reduce the summation in the $B_1$ formula (\ref{B1}) from $N-1$ terms to two terms (for indices $j=N-2$ and $N-1$) only.

The $\widetilde{P}_\Lambda$ term in (\ref{contri1}) can be further calculated through Taylor expansions, and we get
\begin{eqnarray}
&& \widetilde{P}_\Lambda \left[z_1+a_2 R^{-2}, \hspace{0.06cm} z_2 +(k-2\textrm{i}t)R^{-1}, \hspace{0.05cm}  \kappa_1+\frac{1}{6}z_2 R^{-2}, \hspace{0.06cm}  \kappa_2+\frac{1}{12}z_1 R^{-2}, \cdots\right]  \nonumber \\
&& =\widetilde{P}_\Lambda +\frac{\partial \widetilde{P}_\Lambda}{\partial z_2}(k-2\textrm{i}t)R^{-1}+\frac{1}{2}\frac{\partial^2 \widetilde{P}_\Lambda}{\partial z_2^2}(k-2\textrm{i}t)^2R^{-2}+
\frac{\partial \widetilde{P}_\Lambda}{\partial z_1}a_2 R^{-2} + \frac{\partial \widetilde{P}_\Lambda}{\partial \kappa_1}\frac{1}{6}z_2 R^{-2} +\frac{\partial \widetilde{P}_\Lambda}{\partial \kappa_2}\frac{1}{12}z_1 R^{-2}+O(R^{-3}),   \hspace{1cm}   \label{Ptilde}
\end{eqnarray}
where $\widetilde{P}_\Lambda$ and its partial derivatives are evaluated at $(z_1, z_2, \kappa_1, \kappa_2, \dots)$. Notice that $\widetilde{P}_\Lambda$ and its partial derivatives with respect to $z_1$ and $z_2$, when evaluated at $(z_1, z_2, \kappa_1, \kappa_2, \dots)$, are equal to $P_\Lambda$ and its partial derivatives with respect to $z_1$ and $z_2$, because these manipulations do not involve $(\kappa_1, \kappa_2, \dots)$. Putting these results into (\ref{contri1}), we get
\[ \label{LapSplus}
\hspace{-1.0 cm} \det_{1 \leq i, j\leq N} \left(\frac{1}{2^{\nu_j}} S_{n_i-\nu_j}({\textbf{\emph{x}}}^{+}(k)+\nu_j \textbf{\emph{s}}) \right) =2^{-N(N-1)/2}R^\beta \left[
P_\Lambda(z_1, z_2) + \frac{\partial P_\Lambda(z_1, z_2)}{\partial z_2} (k-2\textrm{i}t)R^{-1} + F(z_1, z_2, t, k)R^{-2}+O(R^{-3})\right],
\]
where
\[ \label{defF}
F(z_1, z_2, t, k)\equiv \frac{1}{2}\frac{\partial^2 P_\Lambda}{\partial z_2^2}(k-2\textrm{i}t)^2+
\frac{\partial P_\Lambda}{\partial z_1}a_2 + \frac{\partial \widetilde{P}_\Lambda}{\partial \kappa_1}\frac{1}{6}z_2
+\frac{\partial \widetilde{P}_\Lambda}{\partial \kappa_2}\frac{1}{12}z_1 + s_2 B_1(z_1, z_2).
\]
In this $F$ function, $P_\Lambda$'s derivatives are evaluated at $(z_1, z_2)$ and $\widetilde{P}_\Lambda$'s derivatives evaluated at $(z_1, z_2, \kappa_1, \kappa_2, \dots)$. Notice that when $(\kappa_1, \kappa_2, \dots)$ are treated as variables in the $H_n$ function of (\ref{AMthetak}), we can see that $\partial H_n/\partial \kappa_1=H_{n-3}$ and $\partial H_n/\partial \kappa_2=H_{n-4}$. Utilizing these relations, we can write out the two derivatives of $\widetilde{P}_\Lambda$ in $F$ more explicitly as
\[ \label{Pkappa1}
\frac{\partial \widetilde{P}_\Lambda}{\partial \kappa_1}=\sum_{j=1}^{N}  \det_{1\le i\le N}
         (H_{n_i}, H_{n_i-1}, \cdots,  H_{n_i-(j-2)}, H_{n_i-(j-1)-3},  H_{n_i-j}, \cdots,   H_{n_i-N+1}),
\]
and
\[ \label{Pkappa2}
\frac{\partial \widetilde{P}_\Lambda}{\partial \kappa_2}=\sum_{j=1}^{N}  \det_{1\le i\le N}
         (H_{n_i}, H_{n_i-1}, \cdots,  H_{n_i-(j-2)}, H_{n_i-(j-1)-4},  H_{n_i-j}, \cdots,   H_{n_i-N+1}),
\]
where $H_n=H_n(z_1, z_2)$, and the determinants in (\ref{Pkappa1}) and (\ref{Pkappa2}) are those in (\ref{DoubleRealPolydef}) with sub-indices of their $j$-th column reduced by 3 and 4 respectively.

Similar calculations can be performed for the $S_{n_i-\nu_j} ({\textbf{\emph{x}}}^{-}(k)+ \nu_j \textbf{\emph{s}})$ determinant in the Laplace expansion (\ref{sigmanLap}), and the asymptotic formula for that determinant is that on the right side of Eq.~(\ref{LapSplus}), except that $k-2\textrm{i}t$ should be changed to $-k+2\textrm{i}t$ and
$F(z_1, z_2, t, k)$ changed to $F^*(z_1, z_2, t, -k)$.

\vspace{0.2cm}
(2) Contributions from the index choice of $\boldsymbol{\nu}=(0, 1, \dots, N-2, N)$

For this index choice, utilizing Eq.~(\ref{Snasym}) we see that the $S_{n_i-\nu_j} ({\textbf{\emph{x}}}^{+}(k)+ \nu_j \textbf{\emph{s}})$ determinant in the Laplace expansion (\ref{sigmanLap}) is
\begin{eqnarray}
&&  \det_{1 \leq i, j\leq N} \left(\frac{1}{2^{\nu_j}} S_{n_i-\nu_j}({\textbf{\emph{x}}}^{+}(k)+\nu_j \textbf{\emph{s}}) \right) =2^{-N(N-1)/2-1}R^{\beta-1} \left[\det_{1\le i\le N}\left(S_{n_i}, S_{n_i-1}, \cdots, S_{n_i-(N-2)}, S_{n_i-N}\right)+O(R^{-2})\right], \nonumber \\
&& \label{contri2}
\end{eqnarray}
where $S_n$ on the right side of the above equation is equal to
$S_n [ z_2 +(k-2\textrm{i}t)R^{-1}, \hspace{0.05cm} z_1, \hspace{0.06cm} \kappa_1, \hspace{0.06cm}  \kappa_2, \cdots ]$, which comes from $S_n$'s arguments in Eq.~(\ref{Snasym}) with $O(R^{-2})$ and smaller terms neglected. For these $S_n$ functions, $\partial S_n/\partial z_2=S_{n-1}$. Thus, the determinant on the right side of Eq.~(\ref{contri2}) can be rewritten as $(\partial/\partial z_2) \det_{1\le i\le N}\left(S_{n_i}, S_{n_i-1}, \cdots, S_{n_i-(N-2)}, S_{n_i-(N-1)}\right)$, which is equal to $(\partial/\partial z_2) P_\Lambda[z_1, z_2+(k-2\textrm{i}t)R^{-1}]$. Inserting this result into (\ref{contri2}) and performing Taylor expansion, we get
\begin{eqnarray}
&& \det_{1 \leq i, j\leq N} \left(\frac{1}{2^{\nu_j}} S_{n_i-\nu_j}({\textbf{\emph{x}}}^{+}(k)+\nu_j \textbf{\emph{s}}) \right) =2^{-N(N-1)/2-1}R^{\beta-1} \left[ \frac{\partial P_\Lambda(z_1, z_2)}{\partial z_2}+
\frac{\partial^2 P_\Lambda(z_1, z_2)}{\partial z_2^2} (k-2\textrm{i}t)R^{-1}+O(R^{-2})\right]. \nonumber \\
&&  \label{contri2b}
\end{eqnarray}

Performing similar calculations to the $S_{n_i-\nu_j} ({\textbf{\emph{x}}}^{-}(k)+ \nu_j \textbf{\emph{s}})$ determinant in the Laplace expansion (\ref{sigmanLap}), the asymptotic formula for that determinant is that on the right side of Eq.~(\ref{contri2b}), except that $k-2\textrm{i}t$ should be changed to $-k+2\textrm{i}t$.

\vspace{0.2cm}
(3) Contributions from the index choice of $\boldsymbol{\nu}=(0, 1, \dots, N-2, N+1)$

For this index choice, utilizing Eq.~(\ref{Snasym})  and the fact of $S_{n}(z_2, z_1, \kappa_1, \kappa_2, \dots)=H_{n}(z_1, z_2)$, we get  the $S_{n_i-\nu_j} ({\textbf{\emph{x}}}^{+}(k)+ \nu_j \textbf{\emph{s}})$ determinant in the Laplace expansion (\ref{sigmanLap}) as
\begin{eqnarray}
&& \det_{1 \leq i, j\leq N} \left(\frac{1}{2^{\nu_j}} S_{n_i-\nu_j}({\textbf{\emph{x}}}^{+}(k)+\nu_j \textbf{\emph{s}}) \right) =2^{-N(N-1)/2-2}R^{\beta-2} \left[B_2(z_1, z_2)+O(R^{-1})\right],  \label{contri3}
\end{eqnarray}
where
\[ \label{defB2}
B_2(z_1, z_2)=\det_{1\le i\le N}\left(H_{n_i}, H_{n_i-1}, \cdots, H_{n_i-(N-2)}, H_{n_i-(N+1)}\right),
\]
and $H_n=H_n(z_1, z_2)$. The asymptotic formula for the $S_{n_i-\nu_j} ({\textbf{\emph{x}}}^{-}(k)+ \nu_j \textbf{\emph{s}})$ determinant in that Laplace expansion is the same as the right side of the above equation.

\vspace{0.2cm}
(4) Contributions from the index choice of $\boldsymbol{\nu}=(0, 1, \dots, N-3, N-1, N)$

For this index choice, utilizing Eq.~(\ref{Snasym}) we get  the $S_{n_i-\nu_j} ({\textbf{\emph{x}}}^{+}(k)+ \nu_j \textbf{\emph{s}})$ determinant in the Laplace expansion (\ref{sigmanLap}) as
\begin{eqnarray}
&& \det_{1 \leq i, j\leq N} \left(\frac{1}{2^{\nu_j}} S_{n_i-\nu_j}({\textbf{\emph{x}}}^{+}(k)+\nu_j \textbf{\emph{s}}) \right) =2^{-N(N-1)/2-2}R^{\beta-2} \left[B_3(z_1, z_2)+O(R^{-1})\right],  \hspace{1cm} \label{contri4}
\end{eqnarray}
where
\[ \label{defB3}
B_3(z_1, z_2)=\det_{1\le i\le N}\left(H_{n_i}, H_{n_i-1}, \cdots, H_{n_i-(N-3)}, H_{n_i-(N-1)}, H_{n_i-N}\right),
\]
and $H_n=H_n(z_1, z_2)$. The asymptotic formula for the $S_{n_i-\nu_j} ({\textbf{\emph{x}}}^{-}(k)+ \nu_j \textbf{\emph{s}})$ determinant in that Laplace expansion is the same as the right side of the above equation.

Collecting the above four main contributions to the $\tau_k$ function in the Laplace expansion (\ref{sigmanLap}) and removing an overall constant factor of $2^{-N(N-1)}R^{2\beta}$ that does not affect the solution $(A, Q)$, our spatially uniformly-valid asymptotic approximation for this $\tau_k$ function of the rational solution (\ref{DSAQ}) under multiple large internal parameters (\ref{acond}) and $a_2, t$ being $O(1)$ is
\begin{eqnarray}
&& \tau_k=\left[
P_\Lambda(z_1, z_2) + \frac{\partial P_\Lambda(z_1, z_2)}{\partial z_2} (k-2\textrm{i}t)R^{-1} + F(z_1, z_2, t, k)R^{-2}+O(R^{-3})\right]   \nonumber \\
&&\hspace{0.5cm} \times \left[
P_\Lambda(z_1, z_2) + \frac{\partial P_\Lambda(z_1, z_2)}{\partial z_2} (-k+2\textrm{i}t)R^{-1}
+ F^*(z_1, z_2, t, -k)R^{-2}+O(R^{-3})\right]  \nonumber \\
&&\hspace{0.5cm} +\frac{1}{4}\left[ \frac{\partial P_\Lambda(z_1, z_2)}{\partial z_2}R^{-1}+
\frac{\partial^2 P_\Lambda(z_1, z_2)}{\partial z_2^2} (k-2\textrm{i}t)R^{-2}+O(R^{-3})\right]  \nonumber \\
&&\hspace{0.78cm} \times \left[ \frac{\partial P_\Lambda(z_1, z_2)}{\partial z_2}R^{-1}+
\frac{\partial^2 P_\Lambda(z_1, z_2)}{\partial z_2^2} (-k+2\textrm{i}t)R^{-2}+O(R^{-3})\right] \nonumber \\
&&\hspace{0.5cm} + \frac{1}{16}\left[B_2^2(z_1, z_2)+B_3^2(z_1, z_2)\right]R^{-4}+O(R^{-5}).   \label{taukred}
\end{eqnarray}
Here, the $F$ function is defined in Eq.~(\ref{defF}), and $B_2, B_3$ functions defined in Eqs.~(\ref{defB2}) and (\ref{defB3}).

Compared to the original $\tau_k$ function in Eq.~(\ref{deftaunk}), this reduced $\tau_k$ function in (\ref{taukred}) is simpler for two reasons. First, its time dependence is only quartic, which is much simpler than the time dependence of degree $t^{2\beta}$ in the original $\tau_k$ function ($\beta$ here is given in Eq.~(\ref{defbeta}), which can be very large for a higher-order rational solution). Second, its spatial $(x, y)$ dependence is reflected only through the variables $(z_1, z_2)$ with the connection of $x = 2 z_1 R^2$ and $y = z_2 R$ (see Eq.~(\ref{xyz1z2b})), and this dependence on $(z_1, z_2)$ is in the simple form of certain polynomials and their derivatives.

Next, we graphically show that the above reduced $\tau_k$ function (\ref{taukred}), with terms of $O(R^{-3})$ and $O(R^{-5})$ dropped, leads to approximate DSI solutions that are spatially uniformly valid under parameter conditions (\ref{acond}) and $a_2, t$ being $O(1)$. We use two examples to demonstrate.

\begin{figure}[htb]
\begin{center}
\includegraphics[scale=0.27, bb=1440 0 300 570]{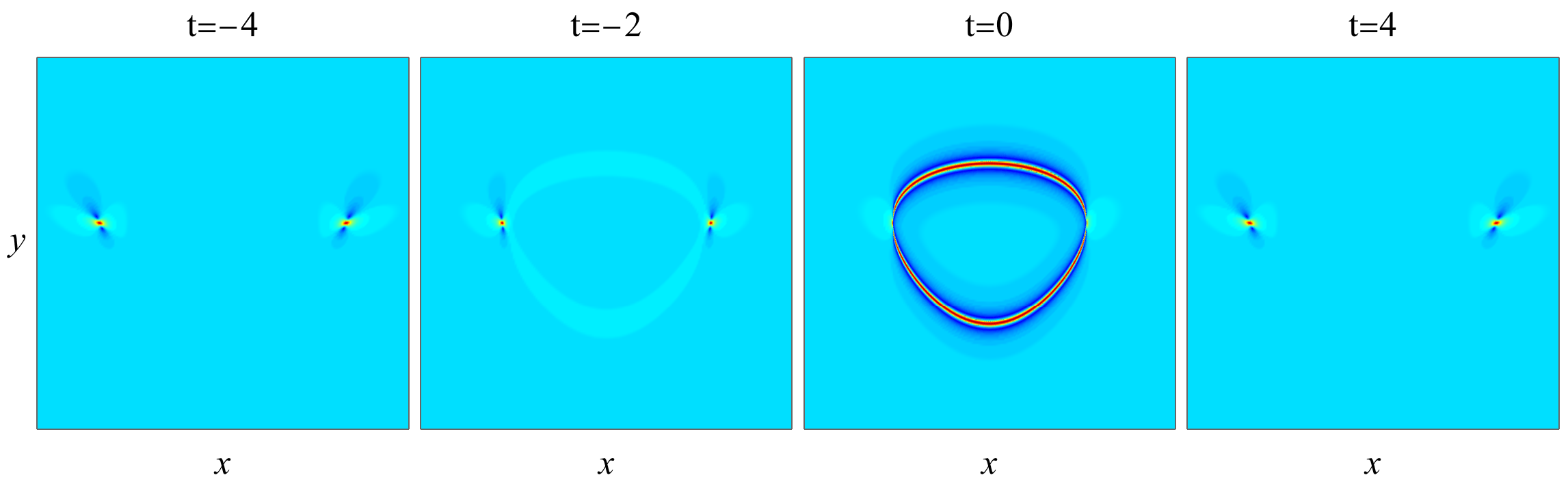}

\includegraphics[scale=0.27, bb=1440 0 300 560]{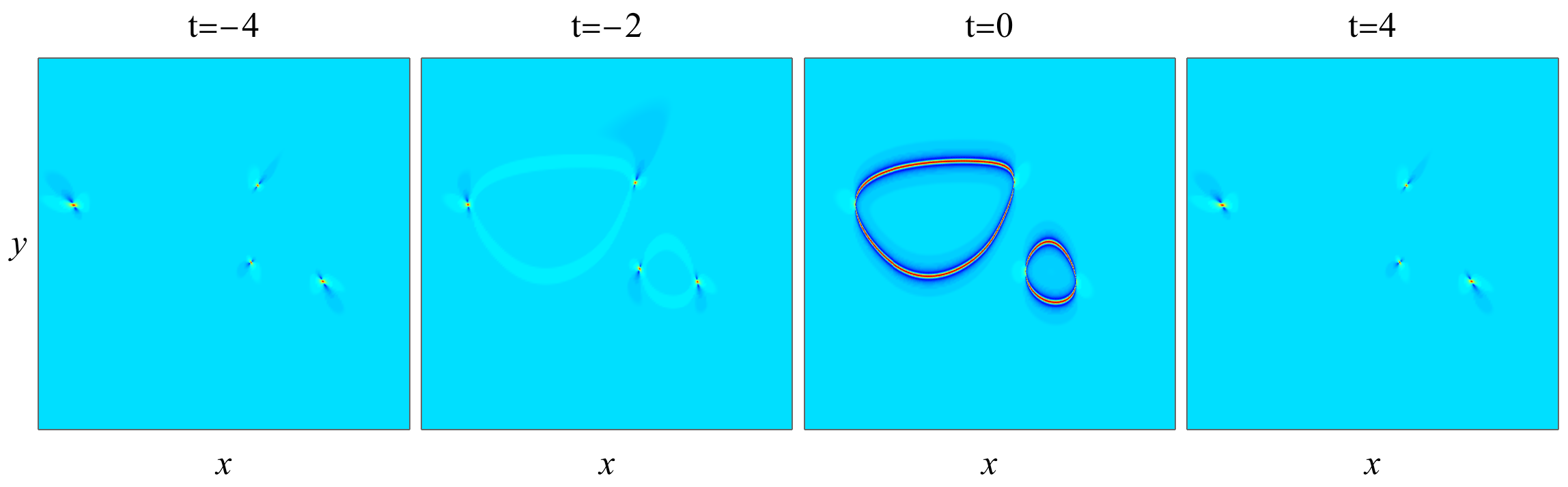}
\caption{Approximate rational solutions $|A(x, y, t)|$ from the spatially uniformly-valid asymptotic formula (\ref{taukred}). Upper row: for the solution shown in Fig.~\ref{f:fig2}; lower row: for the solution shown in the upper row of Fig.~\ref{f:fig4}.  The $(x,y)$ intervals are the same as those in Fig.~\ref{f:fig2} and Fig.~\ref{f:fig4}'s upper row.  } \label{f:fig7}
\end{center}
\end{figure}

One example is the simplest spatially-bounded rogue wave shown in Fig.~\ref{f:fig2}. In this solution, $N=2$, $\Lambda=(2, 3)$, $a_2=0$ and $a_3=1000$. If we put this $a_3$ in the form of $\kappa_1 R^3$ with $R=10$ and $\kappa_1=1$, then the corresponding approximate solution $|A(x, y, t)|$ with $A=\sqrt{2}\tau_1/\tau_0$ and $\tau_k$ from Eq.~(\ref{taukred}) at time values of $t=-4, -2, 0$ and 4 are plotted in the upper row of Fig.~\ref{f:fig7}. When comparing this approximate solution's graphs with those of the exact solution in Fig.~\ref{f:fig2}, we see that this approximate solution provides a good description of the exact solution on the entire $(x, y)$ plane. Indeed, at $t=-2$ and 0, the approximate solution matches the true solution very well. At $t=\pm 4$ where the rogue wave disappears and two lumps remain, we see some differences in the orientations of the approximate and true lumps. This difference is due to the fact that our $R$ value of 10 here is not much larger than the time values of $\pm 4$, while our asymptotic approximation (\ref{taukred}) was derived under the assumption of $R\gg 1$ and $t=O(1)$. We have checked that when we choose larger $R$ values, the approximate lumps at $t=\pm 4$ will become closer to the true lumps.

The other example is the solution shown in the upper row of Fig.~\ref{f:fig4}, which was obtained under parameter choices of Eq.~(\ref{paraFig3a}). In this solution, $N=4$, $\Lambda=(2, 3, 4, 5)$, $a_2=0$, $R=8$, and $(\kappa_1, \kappa_2, \kappa_3)=(1, 2, 3)$. The corresponding approximate solution $|A(x, y, t)|$ with $A=\sqrt{2}\tau_1/\tau_0$ and $\tau_k$ from Eq.~(\ref{taukred}) at time values of $t=-4, -2, 0$ and 4 are plotted in the lower row of Fig.~\ref{f:fig7}. When comparing this approximate solution's graphs with those of the exact solution in the upper row of Fig.~\ref{f:fig4}, we see that this approximate solution also provides a good description of the exact solution on the entire $(x, y)$ plane. As in the previous example, at $t=\pm 4$ differences in the orientations of the approximate and true lumps can be seen, which is due to the same reason that the present $R$ value of 8 is not much larger than the time values of $\pm 4$.

The asymptotic approximation (\ref{taukred}) is valid in the entire $(x, y)$ plane. When we focus on specific regions of this spatial plane, we can get further reduced and simpler approximations. For example, in the $(x, y)$ regions where $\mathcal{P}_{\Lambda}(z_1, z_2)\ne 0$, i.e., when $(x, y)$ is not near the critical curve $\mathcal{P}_{\Lambda}(x/2R^{2}, y/R)=0$, then Eq.~(\ref{taukred}) shows that
\[
\tau_k=\mathcal{P}^2_{\Lambda}(z_1, z_2)+O(R^{-2}).
\]
When this expression is inserted into Eq.~(\ref{DSAQ}), we see that
\[
A(x, y, t)=\sqrt{2}+O(R^{-2}).
\]
In other words, the $A$ solution is approximately the constant background when $(x, y)$ is not near the critical curve, and the error of this background approximation is $O(R^{-2})$. Numerically, we have verified this approximation and its error decay rate.

More interesting are the $(x, y)$ regions near the critical curve, where $\mathcal{P}_{\Lambda}(z_1, z_2)$ is approximately zero. This is the region where nontrivial dynamics occurs. We will perform further reductions to Eq.~(\ref{taukred}) in these regions and show that a rogue wave appears there. Profiles of this rogue wave in these regions will also be derived.

\subsection{Profiles of the rational solution near the critical curve} \label{sec:curve}
When $(x, y)$ is near the critical curve $\mathcal{P}_{\Lambda}(x/2R^{2}, y/R)=0$, we denote
\[ \label{xyxhatyhat}
x=2z_{1c}R^2 +\hat{x}, \quad y=z_{2c}R+\hat{y},
\]
where $\mathcal{P}_{\Lambda}(z_{1c}, z_{2c})= 0$, and $(\hat{x}, \hat{y})=O(1)$. Notice that
this $(z_{1c}, z_{2c})$ is a point on the root curve $\mathcal{P}_{\Lambda}(z_1, z_2)=0$, and $(2z_{1c}R^2, z_{2c}R)$ is a point on the critical curve $\mathcal{P}_{\Lambda}(x/2R^{2}, y/R)=0$. In this case,
in view of the connection (\ref{xyz1z2b}) between $(z_1, z_2)$ and $(x, y)$, we have
\[
z_1=z_{1c}+\frac{1}{2}\hat{x} \hspace{0.04cm} R^{-2}, \quad z_2=z_{2c}+\hat{y} \hspace{0.04cm} R^{-1}.
\]
Substituting these $(z_1, z_2)$ expressions into Eq.~(\ref{taukred}), performing Taylor expansions and then removing an overall factor of $R^{-2}$ from $\tau_k$ that does not affect the solution (\ref{DSAQ}), we get
\begin{eqnarray}
&& \tau_k=\left[
\frac{\partial P_\Lambda(z_{1c}, z_{2c})}{\partial z_2} (\hat{y}+k-2\textrm{i}t) + \widehat{F}(\hat{x}, \hat{y}, t, k)R^{-1}+O(R^{-2})\right]   \nonumber \\
&&\hspace{0.5cm} \times \left[
\frac{\partial P_\Lambda(z_{1c}, z_{2c})}{\partial z_2} (\hat{y}-k+2\textrm{i}t)
+ \widehat{F}^*(\hat{x}, \hat{y}, t, -k)R^{-1}+O(R^{-2})\right]  \nonumber \\
&&\hspace{0.5cm} +\frac{1}{4}\left[\frac{\partial P_\Lambda(z_{1c}, z_{2c})}{\partial z_2}+
\frac{\partial^2 P_\Lambda(z_{1c}, z_{2c})}{\partial z_2^2} (\hat{y}+k-2\textrm{i}t)R^{-1}+O(R^{-2})\right]  \nonumber \\
&&\hspace{0.78cm} \times\left[\frac{\partial P_\Lambda(z_{1c}, z_{2c})}{\partial z_2}+
\frac{\partial^2 P_\Lambda(z_{1c}, z_{2c})}{\partial z_2^2} (\hat{y}-k+2\textrm{i}t)R^{-1}+O(R^{-2})\right]  \nonumber \\
&&\hspace{0.5cm} + \frac{1}{16}\left[B_2^2(z_{1c}, z_{2c})+B_3^2(z_{1c}, z_{2c})\right]R^{-2}+O(R^{-3}),  \label{taukred2}
\end{eqnarray}
where
\[ \label{defFhat}
\widehat{F}(\hat{x}, \hat{y}, t, k)\equiv
\frac{1}{2}\frac{\partial^2 P_\Lambda}{\partial z_2^2}(\hat{y}+k-2\textrm{i}t)^2+
\frac{\partial P_\Lambda}{\partial z_1}\left(\frac{1}{2}\hat{x}+a_2\right) + \frac{\partial \widetilde{P}_\Lambda}{\partial \kappa_1}\frac{1}{6}z_{2c}
+\frac{\partial \widetilde{P}_\Lambda}{\partial \kappa_2}\frac{1}{12}z_{1c} + s_2 B_1(z_{1c}, z_{2c}).
\]
In this $\widehat{F}$ function, $P_\Lambda$'s derivatives are evaluated at $(z_{1c}, z_{2c})$ and $\widetilde{P}_\Lambda$'s derivatives evaluated at $(z_{1c}, z_{2c}, \kappa_1, \kappa_2, \dots)$. The above $\tau_k$ approximation is valid when $R\gg 1$ and $(\hat{x}, \hat{y}, t)=O(1)$.

Further reduction of the above approximation can be made depending on whether $(\partial P_\Lambda/\partial z_2) (z_{1c}, z_{2c})$ is zero or not. Before such further reductions, let us first explain where on the critical curve the quantity $(\partial P_\Lambda/\partial z_2) (z_{1c}, z_{2c})$ is zero. As we have explained in \cite{YangCurve2024}, those points are the bifurcation points of the critical curve if this critical curve is viewed as a bifurcation diagram in the $(x, y)$ plane with $x$ as the bifurcation parameter. We call such points as the exceptional points of the critical curve. For example, horizontal left and right local edges of the critical curve are exceptional points because saddle-node bifurcations occur there. If the critical curve is degenerate and contains isolated points, then $\partial P_\Lambda/\partial z_2$ is zero there as well since the surface of the function $P_\Lambda(z_1, z_2)$ tangentially touches the $(z_1, z_2)$ plane there (as such, $\partial P_\Lambda/\partial z_1$ is zero there too).

\subsubsection{Rogue profiles near the critical curve but away from its exceptional points}

If $(\partial P_\Lambda/\partial z_2) (z_{1c}, z_{2c})\ne 0$, then $(x, y)$ is near the critical curve but away from its exceptional points. According to the second condition of Theorem~\ref{Theorem1}, the critical curve is not degenerate or empty. Thus, the critical curve contains curvy parts where points with $(\partial P_\Lambda/\partial z_2) (z_{1c}, z_{2c})\ne 0$ always exist since they cannot be all bifurcation points. In this case, after removing the nonzero $(\partial P_\Lambda/\partial z_2)^2$ factor from Eq.~(\ref{taukred2}) that does not affect the solution (\ref{DSAQ}), this $\tau_k$ function can be reduced to
\[ \label{taukred3}
\tau_k = (\hat{y}-k+2\textrm{i}t)(\hat{y}+k-2\textrm{i}t)+\frac{1}{4} + O(R^{-1}).
\]
Substituting this expression into (\ref{DSAQ}) and recalling the definition of $\hat{y}$ in Eq.~(\ref{xyxhatyhat}), we find that
\[ \label{APere}
A(x, y, t) = \sqrt{2}\left[ 1+ \frac{16\textrm{i}t-4}{4(y-z_{2c}R)^2+16t^2+1}\right]+O(R^{-1}).
\]
This solution is an approximate Peregrine rogue wave (\ref{1strogue}) along the vertical ($y$) direction, with its crest located on the critical curve. This result is a generalization of a similar result reported in \cite{YangCurve2024} where only one of the internal parameters was large.

The above result confirms that in the $O(1)$ neighborhood of a critical curve, a rogue wave indeed arises. This rogue wave has a Peregrine profile along the vertical direction and is centered at each point of the critical curve (except for exceptional points). Since it has a Peregrine profile, it arises from the constant background and reaches a peak amplitude three times the background and then retreats to that same constant background again. This result analytically explains the appearance of a rogue wave as well as its shape in all previous figures (Figs.~\ref{f:fig2}, \ref{f:fig4}-\ref{f:fig6} and upper row of Fig.~\ref{f:fig7}). In particular, various rogue shapes in all those figures are predicted by the underlying critical curve  $\mathcal{P}_{\Lambda}\left(x/2R^{2}, y/R\right)=0$. This result also analytically explains the fact that when a rogue wave contains multiple pieces (such as three disjoint rings in the lower row of Fig.~\ref{f:fig5}), all these pieces would reach peak amplitude at the same time $t=0$, since that is the time the above Peregrine wave (\ref{APere}) reaches peak amplitude. We have numerically compared the predicted rogue shapes as well as their vertical Peregrine profiles with true rational solutions (\ref{DSAQ}) and found good agreement between them. Details of this comparison will not be shown, however, since a similar comparison has been presented in \cite{YangCurve2024} for the single-large-parameter case.

This Peregrine profile breaks down near an exceptional point of the critical curve, where $(\partial P_\Lambda/\partial z_2) (z_{1c}, z_{2c})=0$. In the next subsection, we will focus on exceptional points out of saddle-node bifurcations (when these exceptional points are viewed as bifurcation points of the critical curve with $x$ as the bifurcation parameter). Such exceptional points are horizontal local left or right edges of the critical curve, i.e., the critical curve horizontally turns around there. The reason we focus on such exceptional points is that they are the generic exceptional points that arise in spatially-bounded rogue waves, see Figs.~\ref{f:fig2} and \ref{f:fig4}-\ref{f:fig6}. For example, in each of the two spatially-bounded rogue waves in Fig.~\ref{f:fig4}, there are four horizontal local edges of the critical curve, two on the left and two on the right, and there are no other exceptional points (see Fig.~\ref{f:fig3}(a, b)). Thus, if the solution profile near such edge-type exceptional points is clarified, then we would get a generically complete analytical understanding of spatially-bounded rogue waves in DSI, which is the subject of this paper. Of course, such edge-type exceptional points can also arise in spatially-unbounded rogue waves, see the upper row of Fig.~\ref{f:fig6} and Ref.~\cite{YangCurve2024}.

\subsubsection{Solution profiles near horizontal local edges of the critical curve} \label{sec:edge}

Out of a saddle-node bifurcation, the exceptional point is a horizontal local left or right edge of the critical curve. Since we have shown above that a Peregrine-type rogue wave appears almost everywhere near the critical curve and the rogue crest is on this critical curve, this edge of the critical curve then is also an edge of the rogue crest. For such an edge point, we consider the generic case where $(\partial^2 P_\Lambda/\partial z_2^2) (z_{1c}, z_{2c})\ne 0$. Thus, we have
\[ \label{EdgeCond}
P_\Lambda(z_{1c}, z_{2c})=0, \quad \frac{\partial P_\Lambda (z_{1c}, z_{2c})}{\partial z_2}=0, \quad
\frac{\partial^2 P_\Lambda (z_{1c}, z_{2c})}{\partial z_2^2}\ne0.
\]
Below, we investigate the solution profile near such edges.

Utilizing conditions (\ref{EdgeCond}) and removing a certain constant factor from the $\tau_k$ function in Eq.~(\ref{taukred2}) which does not affect the solution (\ref{DSAQ}), this $\tau_k$ reduces to
\begin{eqnarray}
&& \tau_k=\left[
\frac{1}{2}(\hat{y}+k-2\textrm{i}t)^2+\gamma_1\left(\frac{1}{2}\hat{x}+a_2\right) +\gamma_2 \right]   \left[\frac{1}{2}(\hat{y}-k+2\textrm{i}t)^2+\gamma_1\left(\frac{1}{2}\hat{x}+a_2^*\right) + \gamma_2 \right]
\nonumber \\
&&\hspace{0.5cm} +\frac{1}{4} (\hat{y}+k-2\textrm{i}t)(\hat{y}-k+2\textrm{i}t) + \frac{1}{16}\gamma_3+O(R^{-1}),  \label{taukred4}
\end{eqnarray}
where
\begin{eqnarray} \label{defGamma123}
&& \gamma_1=\left(\frac{\partial^2 P_\Lambda}{\partial z_2^2}\right)^{-1}\frac{\partial P_\Lambda}{\partial z_1}, \quad \gamma_2=\left(\frac{\partial^2 P_\Lambda}{\partial z_2^2}\right)^{-1}\left[\frac{\partial \widetilde{P}_\Lambda}{\partial \kappa_1}\frac{1}{6}z_{2c}
+\frac{\partial \widetilde{P}_\Lambda}{\partial \kappa_2}\frac{1}{12}z_{1c} + s_2 B_1(z_{1c}, z_{2c})\right],   \label{defGamma123a} \\
&&\gamma_3=\left(\frac{\partial^2 P_\Lambda}{\partial z_2^2}\right)^{-2}\left[B_2^2(z_{1c}, z_{2c})+B_3^2(z_{1c}, z_{2c})\right].   \label{defGamma123b}
\end{eqnarray}
In these $\gamma_k$ quantities, $P_\Lambda$'s derivatives are evaluated at $(z_{1c}, z_{2c})$ and $\widetilde{P}_\Lambda$'s derivatives evaluated at $(z_{1c}, z_{2c}, \kappa_1, \kappa_2, \dots)$. Notice that all three $\gamma_k$'s are real-valued. The above $\tau_k$ approximation (\ref{taukred4}) is valid when $R\gg 1$, $(\hat{x}, \hat{y}, t)=O(1)$, and conditions (\ref{EdgeCond}) hold .

In the appendix, we will show that $\gamma_1=\pm 1$, with the plus sign for the right local edge and the minus sign for the left local edge. In addition, we will show in the appendix that $\gamma_3=1$. Thus, at a right edge, our reduced $\tau_k$ function (\ref{taukred4}) is
\[
\tau_k=\left[
\frac{1}{2}(\hat{y}+k-2\textrm{i}t)^2+\frac{1}{2}(\hat{x}+2\gamma_2)+a_2 \right]   \left[\frac{1}{2}(\hat{y}-k+2\textrm{i}t)^2+\frac{1}{2}(\hat{x}+2\gamma_2)+a_2^* \right] +\frac{1}{4} (\hat{y}+k-2\textrm{i}t)(\hat{y}-k+2\textrm{i}t) + \frac{1}{16}+O(R^{-1}),  \label{taukred5a}
\]
and at a left edge, the reduced $\tau_k$ function (\ref{taukred4}) is
\[
\tau_k=\left[
\frac{1}{2}(\hat{y}+k-2\textrm{i}t)^2-(\frac{1}{2}\hat{x}-2\gamma_2)-a_2 \right]   \left[\frac{1}{2}(\hat{y}-k+2\textrm{i}t)^2-\frac{1}{2}(\hat{x}-2\gamma_2)-a_2^* \right] +\frac{1}{4} (\hat{y}+k-2\textrm{i}t)(\hat{y}-k+2\textrm{i}t) + \frac{1}{16}+O(R^{-1}).  \label{taukred5b}
\]
Comparing these two $\tau_k$ functions with the second-order rational solution (\ref{2ndroguetauk}) and its $x$-direction reversal (\ref{2ndroguetauk2}), we see that (\ref{taukred5a}) is just the second-order rational solution (\ref{2ndroguetauk}) with a shifted $(x, y)$ axes, and (\ref{taukred5b}) is just the second-order rational solution (\ref{2ndroguetauk2}) with a shifted $(x, y)$ axes. In other words, the higher-order rational solution (\ref{DSAQ}) near a horizontal local edge of the critical curve (i.e., the rogue crest) is asymptotically a second-order rational solution (\ref{2ndroguetauk}) or (\ref{2ndroguetauk2}).

\begin{figure}[htb]
\begin{center}
\includegraphics[scale=0.27, bb=1440 0 300 570]{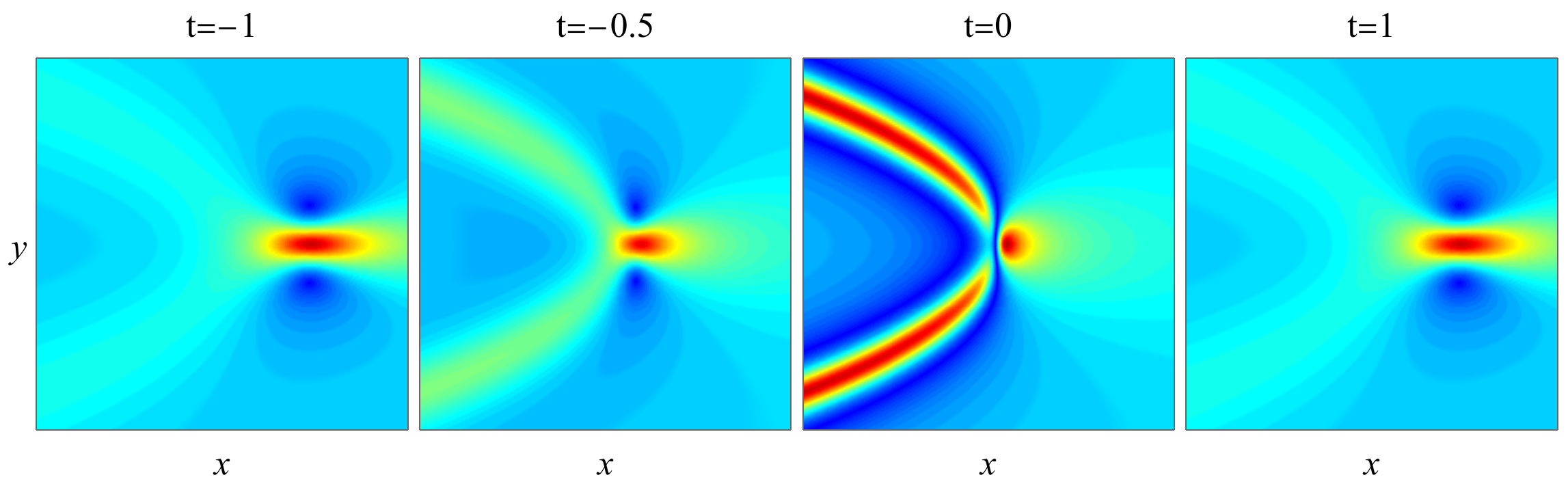}

\includegraphics[scale=0.27, bb=1440 0 300 560]{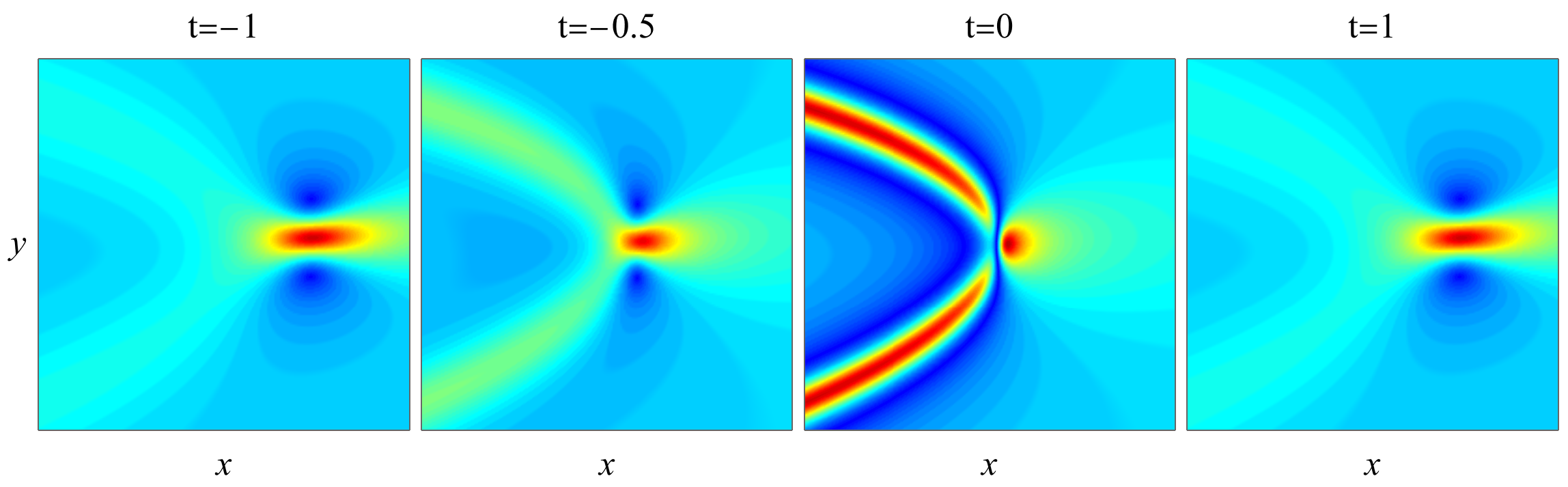}
\caption{Comparison between the true rational solution $|A|$ from Eq.~(\ref{DSAQ}) and its asymptotic approximation (\ref{taukred5a}) near the right edge of the rogue wave in Fig.~\ref{f:fig2} at four time values of $t=-1, -0.5, 0$ and 1. Upper row: the asymptotic approximation (\ref{taukred5a}); lower row: the true solution. In all panels, the $(x, y)$ intervals are $|\hat{x}|\le 10$ and $|\hat{y}|\le 4$, i.e., $|x-2z_{1c}R^2|\le 10$ and $|y-z_{2c}R|\le 4$, where $(2z_{1c}R^2, z_{2c}R)$ is the right edge of the critical curve with $(z_{1c}, z_{2c})=(3^{2/3}/2, 3^{1/3})$ and $R=10$.  } \label{f:fig8}
\end{center}
\end{figure}

Next, we use an example to compare these asymptotic predictions (\ref{taukred5a})-(\ref{taukred5b}) with true rational solutions (\ref{DSAQ}) near horizontal local edges of their rogue crests. For simplicity, we use the simplest spatially-bounded rogue wave (\ref{simpleLRogue}) shown in Fig.~\ref{f:fig2} as the example. In this example, $\Lambda=(2, 3)$, $a_2=0$ and $a_3=1000$. We put this $a_3$ in the form of $\kappa_1 R^3$ with $R=10$ and $\kappa_1=1$. Then, we find that
\[
P_\Lambda (z_{1}, z_{2})=z_1^2-z_2+\frac{1}{12}z_2^4.
\]
The graph of its root curve is a single ring whose shape resembles the rogue wave shown in Fig.~\ref{f:fig2}. The right edge of this root curve is at $(z_{1c}, z_{2c})=(3^{2/3}/2, 3^{1/3})$. At this edge, we find that $\gamma_2=-1/4$. Inserting the relation $\hat{x}=x-2z_{1c}R^2$ and $\hat{y}=y-z_{2c}R$ from Eq.~(\ref{xyxhatyhat}) into the $\tau_k$ function in Eq.~(\ref{taukred5a}) and neglecting its $O(R^{-1})$ error term, the resulting $A(x, y, t)=\sqrt{2}\tau_1/\tau_0$ approximation at four time values of $t=-1, -0.5, 0$ and 1 is plotted in the upper row of Fig.~\ref{f:fig8} for the spatial domain of $|\hat{x}|\le 10$ and $|\hat{y}|\le 4$. For comparison, the true solution (\ref{simpleLRogue}) in the same spatial domains and at the same time values is plotted in the lower row of Fig.~\ref{f:fig8}. It is seen that the asymptotic approximation matches the true solution quite well,
even for $|\hat{x}|$ values up to 10, although our asymptotic approximation was derived only for $(\hat{x}, \hat{y})=O(1)$. We have also numerically examined the error of this asymptotic approximation versus the $R$ value at the location $(\hat{x}, \hat{y})=(1, 1)$ and confirmed the $O(R^{-1})$ decay rate predicted in Eq.~(\ref{taukred5a}), and details will be omitted here for brevity. These agreements confirm that at a horizontal local edge of a rogue wave, the higher-order rational solution (\ref{DSAQ}) is indeed an approximate second-order rational solution (\ref{2ndroguetauk}) or (\ref{2ndroguetauk2}). Because of this, rational solutions (\ref{DSAQ}) near all local edges of spatially-bounded rogue waves of Figs.~\ref{f:fig2}, \ref{f:fig4} and \ref{f:fig5} as well as spatially-unbounded rogue waves in the upper row of Fig.~\ref{f:fig6} are approximate second-order rational solutions. As such, all lumps that we see in those figures are the type of lumps in second-order rational solutions (see Fig.~\ref{f:fig1}) that shrink or expand horizontally over time, not the type of shape-preserving lumps moving on a constant background that the DSI equation also admits \cite{Satsuma_Ablowitz,OhtaYangDSI}.

\section{Summary and discussion}
In summary, we have determined spatially-bounded rogue waves in the Davey-Stewartson I equation. We have shown that such rogue waves can be obtained when a single or multiple internal parameters in the higher-order rational solution (\ref{DSAQ}) are real and large if two conditions are met. One condition is that the order-index vector of this higher-order rational solution has even length and comprises pairs of the form $(2n, 2n+1)$, where $n$ is a positive integer. The other condition is that the root curve of the associated double-real-variable polynomial equation (\ref{Pmz1z2}) is not degenerate or empty. Under these conditions, spatially-bounded rogue waves will arise from a uniform background with some time-varying lumps on it, reach high amplitude in limited space, and then disappear into the same background again. The crests of these rogue waves form a single or multiple closed curves that are generically disconnected from each other on the spatial plane, and these crests are asymptotically predicted by the root curve of the polynomial equation (\ref{Pmz1z2}). Analytically, we have derived uniformly-valid asymptotic approximations for these spatially-bounded rogue waves in the large-parameter regime. Near the crests of these rogue waves, these asymptotic approximations reduce to simple expressions. Specifically, we have shown that a higher-order rational solution near a horizontal local edge of its rogue crest is asymptotically a simple second-order rational solution which contains a rogue parabola and a time-varying lump, while this solution between horizontal edges has a vertical Peregrine rogue profile. Our asymptotic approximations of these rogue waves have been compared to true solutions and good agreement has been demonstrated.

One notable feature about rogue waves in this paper is that, when they contain multiple pieces, these different rogue pieces would appear and disappear at the same time, not one after another. This feature can be graphically seen in Figs.~\ref{f:fig4}-\ref{f:fig6} and analytically explained by our Peregrine approximation (\ref{APere}). An interesting question is whether the higher-order rational solutions (\ref{DSAQ}) admit rogue waves whose different parts appear and disappear sequentially rather than simultaneously. This question may be studied in parameter regimes different from (\ref{acond}) of this paper and will be left for future studies.

Another interesting question is the extension of these results to other multidimensional integrable equations. For example, whether spatially-bounded rogue waves can be obtained in (2+1)-dimensional three-wave resonant interaction system and others remains to be seen. These extensions fall outside the scope of this paper and will be left for future studies as well.

\section*{Acknowledgment}
The work of B.Y. was supported in part by the National Natural Science Foundation of China (Grant No.12201326, 12431008).

\section*{Appendix}
\renewcommand{\theequation}{A.\arabic{equation}}

In this appendix, we prove that $\gamma_1=\pm 1$ and $\gamma_3=1$, where $\gamma_1$ and $\gamma_3$ are defined in Eqs.~(\ref{defGamma123a})-(\ref{defGamma123b}).

First, we show $\gamma_1=\pm 1$. For the $P_\Lambda(z_1, z_2)$ function defined in Eq.~(\ref{DoubleRealPolydef}), using the relations of $(\partial/\partial z_2)H_{n}(z_1, z_2)=H_{n-1}(z_1, z_2)$ and $(\partial/\partial z_1)H_{n}(z_1, z_2)=H_{n-2}(z_1, z_2)$, we find that
\[ \label{Pz1Pz2z2}
\frac{\partial P_\Lambda}{\partial z_1}=c_1-c_2, \quad \frac{\partial^2 P_\Lambda}{\partial z_2^2}=c_1+c_2,
\]
where
\begin{eqnarray*}
&& c_1=\det_{1\le i\le N}\left(H_{n_i}, H_{n_i-1}, \cdots, H_{n_i-(N-3)}, H_{n_i-(N-2)}, H_{n_i-(N+1)}\right), \\
&& c_2=\det_{1\le i\le N}\left(H_{n_i}, H_{n_i-1}, \cdots, H_{n_i-(N-3)}, H_{n_i-(N-1)}, H_{n_i-N}\right).
\end{eqnarray*}
We additionally denote
\begin{eqnarray*}
&& c_3=\det_{1\le i\le N}\left(H_{n_i}, H_{n_i-1}, \cdots, H_{n_i-(N-3)}, H_{n_i-(N-2)}, H_{n_i-N}\right), \\
&& c_4=\det_{1\le i\le N}\left(H_{n_i}, H_{n_i-1}, \cdots, H_{n_i-(N-3)}, H_{n_i-(N-1)}, H_{n_i-(N+1)}\right), \\
&& c_5=\det_{1\le i\le N}\left(H_{n_i}, H_{n_i-1}, \cdots, H_{n_i-(N-3)}, H_{n_i-(N-2)}, H_{n_i-(N-1)}\right), \\
&& c_6=\det_{1\le i\le N}\left(H_{n_i}, H_{n_i-1}, \cdots, H_{n_i-(N-3)}, H_{n_i-N}, H_{n_i-(N+1)}\right).
\end{eqnarray*}
Notice that only the last two columns of these six determinants are different. In addition, those last two columns are six combinations of the four vectors of $H_{n_i-(N-2)}$, $H_{n_i-(N-1)}$, $H_{n_i-N}$ and $H_{n_i-(N+1)}$. Then,  the Plucker relation between these six determinants gives
\[
c_1c_2-c_3c_4+c_5c_6=0.
\]
Notice also that $c_5=P_\Lambda(z_1, z_2)$ and $c_3=\partial P_\Lambda(z_1, z_2)/\partial z_2$. At an exceptional point $(z_{1c}, z_{2c})$, $c_3=c_5=0$ (see Eq.~(\ref{EdgeCond})). Thus, the above Plucker relation reduces to
\[ \label{b1b2zero}
c_1c_2=0.
\]
Under our conditions (\ref{EdgeCond}) for horizontal edge points of the critical curve, $\partial^2 P_\Lambda/\partial z_2^2\ne 0$, i.e., $c_1+c_2\ne 0$. Combining this result with (\ref{b1b2zero}), we conclude that only one of $c_1$ and $c_2$ is zero. In this case, Eqs.~(\ref{Pz1Pz2z2}) and (\ref{defGamma123a}) then show that $\gamma_1=\pm 1$.

To determine whether $\gamma_1=1$ or $-1$ at an edge, let us view the root curve near that edge as $z_1$ being a single-valued function of $z_2$ (since the root curve shows a saddle-node bifurcation there). Then, when we differentiate the equation $P_\Lambda(z_1, z_2)=0$ with respect to $z_2$ twice, we get
\[
\frac{\partial^2 P_\Lambda}{\partial z_2^2}+\frac{\partial^2 P_\Lambda}{\partial z_1 \partial z_2} \frac{dz_1}{dz_2}+
\left( \frac{\partial^2 P_\Lambda}{\partial z_1 \partial z_2} +\frac{\partial^2 P_\Lambda}{\partial z_1^2}\frac{dz_1}{dz_2}\right)\frac{dz_1}{dz_2}+
\frac{\partial P_\Lambda}{\partial z_1}  \frac{d^2z_1}{dz_2^2}=0.
\]
At an edge point $(z_{1c}, z_{2c})$, $dz_1/dz_2=0$. Then, recalling the definition of $\gamma_1$ in Eq.~(\ref{defGamma123a}), the above equation reduces to
\[
\gamma_1\frac{d^2z_1}{dz_2^2}=-1.
\]
Using the fact of $\gamma_1=\pm 1$ obtained above, we see that $d^2z_1/dz_2^2=\pm 1$ at a horizontal local edge point. Since $d^2z_1/dz_2^2$ is negative at a right edge and positive at a left edge, we then see that $\gamma_1=1$ at a right edge and $\gamma_1=-1$ at a left edge.

Next, we show $\gamma_3=1$. From the definitions of $B_2$ and $B_3$ in Eqs.~(\ref{defB2}) and (\ref{defB3}), we see that $B_2=c_1$ and $B_3=c_2$, where $c_1$ and $c_2$ are given earlier in this appendix.
At a horizontal edge point, since one and only one of $c_1$ and $c_2$ is zero, we have
\[
B_2^2+B_3^2=c_1^2+c_2^2=(c_1+c_2)^2=\left(\frac{\partial^2 P_\Lambda}{\partial z_2^2}\right)^{2}.
\]
Thus, $\gamma_3=1$ from its definition in Eq.~(\ref{defGamma123b}).

\end{document}